\documentclass[aps,pra,twocolumn,superscriptaddress,showpacs]{revtex4-1}
\usepackage{graphicx}
\usepackage{subfigure}
\usepackage{amsmath}
\def\gappeq{\mathrel{ \rlap{\raise.5ex\hbox{$>$}}
                      {\lower.5ex\hbox{$\sim$}} } }
\def\lappeq{\mathrel{ \rlap{\raise.5ex\hbox{$<$}}
                      {\lower.5ex\hbox{$\sim$}} } }
\newcommand{\pa}{\partial}

\begin{document}

\title{Equilibrium solutions of immiscible two-species Bose-Einstein condensates in perturbed harmonic traps}

\author{R. W. Pattinson}
\email[]{r.w.pattinson@ncl.ac.uk}
\affiliation{Joint Quantum Centre Durham--Newcastle, School of Mathematics and Statistics, Newcastle University, Newcastle upon Tyne, NE1 7RU, United Kingdom}

\author{T. P. Billam}
\affiliation{Jack Dodd Centre for Quantum Technology, Department of Physics, University of Otago, Dunedin, 9016, New Zealand}

\author{S. A. Gardiner}
\affiliation{Joint Quantum Centre Durham--Newcastle, Department of Physics, Durham University, Durham, DH1 3LE, United Kingdom}
\author{D. J. McCarron}
\affiliation{Joint Quantum Centre Durham--Newcastle, Department of Physics, Durham University, Durham, DH1 3LE, United Kingdom}
\author{H. W. Cho}
\affiliation{Joint Quantum Centre Durham--Newcastle, Department of Physics, Durham University, Durham, DH1 3LE, United Kingdom}
\author{S. L. Cornish}
\affiliation{Joint Quantum Centre Durham--Newcastle, Department of Physics, Durham University, Durham, DH1 3LE, United Kingdom}

\author{N. G. Parker}
\affiliation{Joint Quantum Centre Durham--Newcastle, School of Mathematics and Statistics, Newcastle University, Newcastle upon Tyne, NE1 7RU, United Kingdom}
\author{N. P. Proukakis}
\affiliation{Joint Quantum Centre Durham--Newcastle, School of Mathematics and Statistics, Newcastle University, Newcastle upon Tyne, NE1 7RU, United Kingdom}

\date{\today}

\begin{abstract}

We investigate the mean--field equilibrium solutions for a two--species immiscible Bose--Einstein condensate confined by a harmonic confinement with additional linear perturbations. We observe a range of equilibrium density structures, including `ball and shell' formations and axially/radially separated states, with a marked sensitivity to the potential perturbations and the relative atom number in each species. Incorporation of linear trap perturbations, albeit weak, are found to be essential to match the range of equilibrium density profiles observed in a recent $^{87}$Rb--$^{133}$Cs Bose–Einstein condensate experiment [D. J. McCarron et al., Phys. Rev. A, 84, 011603(R) (2011)]. Our analysis of this experiment demonstrates that sensitivity to linear trap perturbations is likely to be important factor in interpreting the results of similar experiments in the future.

\end{abstract}

%\keywords{}

\maketitle


\section{Introduction\label{intro}}

  Since the successful realization of an atomic Bose--Einstein condensate (BEC) composed of two different hyperfine spin states of $^{87}$Rb~\cite{PhysRevLett.78.586}, experimental and theoretical work has advanced greatly in the field of two--component BECs. These have been produced using different atomic species~\cite{PhysRevLett.89.053202,PhysRevLett.89.190404,PhysRevLett.100.210402,Durham}, different isotopes of the same atom~\cite{PhysRevLett.101.040402}, and a single isotope in two different hyperfine spin states~\cite{PhysRevLett.78.586,PhysRevLett.81.1539,PhysRevLett.85.2413,PhysRevA.63.051602,PhysRevLett.99.190402,PhysRevA.80.023603,PhysRevA.82.033609}.  Spinor condensates, which have at least three components with internal spin degrees of freedom, are also generating much current interest (see~\cite{spinor.rev} for a review).  A key feature of two--species BECs is their potential to exhibit miscible or immiscible behaviour depending on the inter-species interactions. Immiscibility, where repulsion between species favours their spatial separation, has been observed~\cite{Durham,PhysRevLett.101.040402,PhysRevLett.81.1539,PhysRevA.82.033609}. In recent years, many static and dynamical properties of two-species BECs have been analysed. These include ground state structures~\cite{PhysRevLett.77.3276,PhysRevLett.80.1130,PhysRevLett.81.5718,PhysRevA.58.4836,Tripp,PhysRevA.66.013612,PhysRevA.78.023624,0953-4075-43-9-095302, PhysRevA.66.015602, PhysRevA.58.1440, PhysRevA.62.053601, PhysRevE.65.066201,PhysRevA.85.023613}, modulation instabilities~\cite{PhysRevA.64.021601,Shukla_Sten_Fedele_2001,PhysRevLett.93.100402,Kourakis_Shukla_Marklund_Stenflo_2005,PhysRevA.71.035601,modulation}, dark--bright solitons~\cite{PhysRevLett.87.010401,1367-2630-5-1-364,Becker_Stellmer_Soltan-Panahi_Drscher_Baumert_Richter_Kronjger_Bongs_Sengstock_2008,PhysRevLett.106.065302}, vortices~\cite{PhysRevLett.83.2498,PhysRevLett.82.4956}, and the role of finite temperature~\cite{PhysRevLett.78.3594,PhysRevA.57.1272,PhysRevA.75.013601,PhysRevA.77.033606, PhysRevA.70.063606}.   In the limit of zero temperature, the mean-field of a single or two-component condensate is described by the Gross-Pitaevskii equation, in either single or coupled form, respectively.  For immiscible two-component condensates under cylindrically symmetric trapping, the mean-field ground state has been shown to exist in a phase separated structure~\cite{PhysRevLett.77.3276, PhysRevLett.80.1130} where one component lies at the trap centre with the other lying at the periphery. This symmetry can be broken to give rise to two separated side--by--side condensates~\cite{PhysRevA.58.1440, Tripp,0953-4075-43-9-095302, PhysRevA.66.015602,PhysRevA.85.023613}. 

The aim of this paper is to study how relatively minor experimentally-relevant asymmetric trap perturbations can modify the equilibrium density structures that arise in an otherwise harmonically trapped immiscible two--species BEC. Under the conditions probed, the resulting structures are expected to be dominated by the presence of the condensates appearing in each component, and so we perform our analysis using zero temperature mean--field theory.

Since the trap perturbations present in any particular experiment will vary with technical details, as a case study we focus on the $^{87}$Rb--$^{133}$Cs (referred to as Rb and Cs hereinafter) system, for which a recent experiment~\cite{Durham} revealed three regimes of density structures, depending on the relative atom numbers in each species.  We show that, although under harmonic potentials alone, the equilibrium solutions do not fully match the experimental results, the dramatic effects arising from the incorporation of additional weak linear shifts to the potentials leads to the bulk features of the experimental observations being recovered, even within our simplified model.

In Sec.~\ref{theory}, we briefly review the three density structure regimes observed experimentally and the coupled mean-field Gross--Pitaevskii equations, also presenting the equilibrium density profiles in unperturbed harmonic traps (which fail to match the experimental results). Section~\ref{offset} examines the effects of adding perturbing linear potentials to the harmonic traps, in both the axial and a transverse direction, and demonstrates how this modifies the obtained structures, such that the observed features can be recovered (for suitably-identified experimentally-relevant values of these perturbing potentials). Conclusions and additional relevant remarks are given in Section~\ref{conclude} and an Appendix.

\section{Motivation and Theory\label{theory}}

\subsection{Experimental motivation}

A recent Rb--Cs two-species BEC experiment~\cite{Durham} revealed distinct regimes of density distributions depending on the relative atom numbers in each species. This experimental method relies on sympathetic cooling of Cs atoms via evaporatively-cooled Rb atoms, confined in a levitated crossed dipole trap~\cite{jenkin_2011}. While large inelastic three body losses~\cite{cho_2011} between the species is an obstacle to achieving high phase space densities for this mixture, this was to a degree overcome  by tilting the dipole trap (using an applied magnetic field gradient). Two species condensates were produced with up to $\sim2\times10^{4}$~atoms of each species. The intraspecies and interspecies {\it s}--wave scattering lengths in the experiment were $a_{\textrm{Rb}}=100~a_{0}$~\cite{PhysRevLett.89.283202}, $a_{\textrm{Cs}}=280~a_{0}$~\cite{PhysRevA.70.032701} and $a_{\textrm{RbCs}}=650~a_{0}$~\cite{PhysRevA.85.032506}. In modelling the experiment we describe the optical dipole trap as a cylindrically-symmetric harmonic potential,    
\begin{eqnarray*}
      V_{i}\left(x,y,z\right)=\frac{1}{2}m_{i}\left[\omega_{\perp(i)}^{2}(x^{2}+y^{2})+\omega_{z(i)}^{2}z^{2}\right],
\label{eqn:V}
\end{eqnarray*} 
where $i=\textrm{Rb},~\textrm{Cs}$.
The trap frequencies are $\omega_{\perp(\textrm{Rb})}=2\pi\times32.2~\textrm{Hz}$, $\omega_{\perp(\textrm{Cs})}=2\pi\times40.2~\textrm{Hz}$ in the transverse directions and $\omega_{z(\textrm{Rb})}=2\pi\times3.89~\textrm{Hz}$, $\omega_{z(\textrm{Cs})}=2\pi\times4.55~\textrm{Hz}$ in the axial direction.    

While the trapping is dominantly harmonic according to Eq.~(\ref{eqn:V}), weak perturbations existed in the experiment which must be accounted for. The above-mentioned magnetic tilt, which differs slightly between the species, is applied in one of the transverse directions and results in a shift in relative trap centres by up to 3 microns transversely.  Additionally, the small difference in magnetic moment-to-mass ratio for each species, coupled with minute unavoidable misalignments of the dipole trap beams with respect to the magnetic potential, may result in offsets between the trap centres of up to $2~\mu$m in all directions. Further trap perturbations, such as differential gravitational sag, are also present. 

\begin{figure*}
        \includegraphics[width=1.0\textwidth,clip]{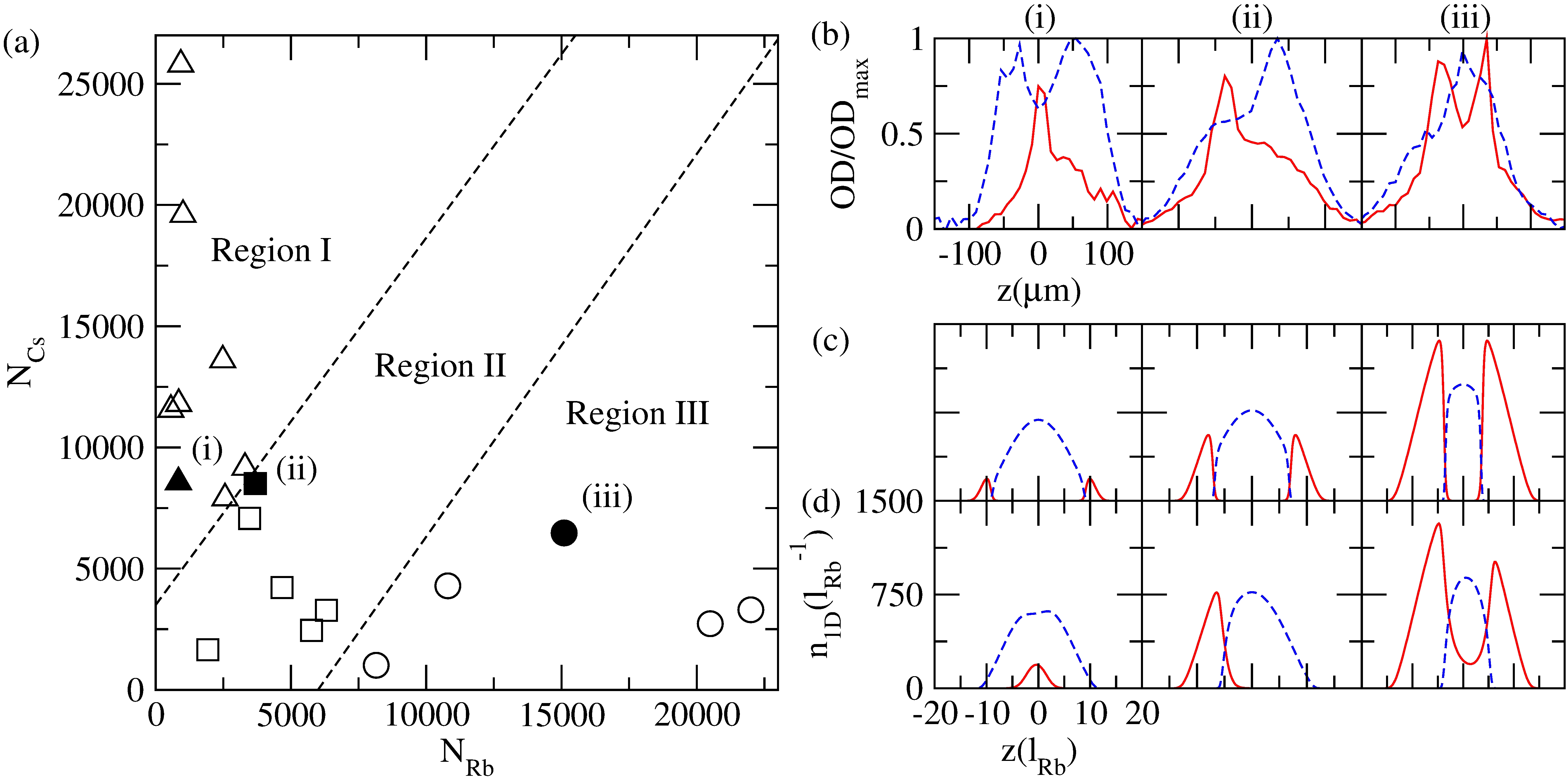}
        \caption{(Color online) (a) Experimental data for the $^{87}$Rb--$^{133}$Cs BEC experiment of Ref. \cite{Durham}. Depending on relative atom numbers, three distinct structures are observed, represented here through triangles, squares and circles (Regions I, II and III). (b) Experimental integrated axial density profiles corresponding to the filled symbols in (a), observed after time-of-flight expansion~\cite{exp_note} and rescaled to the optical depth (OD) maximum. (c) Mean-field cylindrically symmetric ground state density profiles corresponding to the atom numbers in (b). (d) Integrated axial ground state profiles under an axial linear potential $\delta_{z}=0.9~\mu$m and a transverse linear potential $\delta_{x}=1.0~\mu$m. (Solid) red curve --- Rb; (dashed) blue curve --- Cs.\label{fig:points}}
      \end{figure*}

In the experiment, it was observed that the density profiles fell into three distinct regimes depending on relative atom numbers of the species.  This is summarised in Fig.~\ref{fig:points} (a) where the three regimes are labelled Region I (triangles), II (squares) and III (circles).   We consider one representative set of atom numbers from each structural regime:
\begin{itemize}
\item[] (i)~~~$N_{\textrm{Rb}}=840$,~~~$N_{\textrm{Cs}}=8570$
\item[] (ii)~~$N_{\textrm{Rb}}=3680$,~~$N_{\textrm{Cs}}=8510$
\item[] (iii)~$N_{\textrm{Rb}}=15100$,~$N_{\textrm{Cs}}=6470$
\end{itemize}
These test cases are indicated by the filled symbols in Fig.~\ref{fig:points} (a).  The corresponding experimental images of the axial density profile are presented in Fig.~\ref{fig:points} (b) and serve to illustrate the different structures obtained in each regime.  For Regions I and III, one of two possible symmetric cases is obtained: the Rb sits in the centre for Region I while the Rb is spatially split by the Cs in Region III. In Region II, the condensates adopt asymmetric density profiles, sitting side-by-side along the weaker axial direction of the trap. The experimental images have undergone time--of--flight expansion and include broad thermal density profiles, and so our analysis is limited to the qualitative structural form only (discussed in Sec.~\ref{conclude}).

\subsection{Mean-field theory of two-species BECs}
In the limit of near--zero temperatures, the mean-field of a two--species BEC is well-described by a set of coupled Gross--Pitaevskii equations (CGPEs) \cite{PhysRevLett.77.3276},
  \begin{widetext}  
    \begin{equation}
      i\hbar\frac{\pa\psi_{Rb}}{\pa t}=\left(-\frac{\hbar^{2}}{2m_{Rb}}\nabla^{2}+V_{Rb}+g_{Rb}\left|\psi_{Rb}\right|^{2}+g_{RbCs}\left|\psi_{Cs}\right|^{2}-\mu_{Rb}\right)\psi_{Rb}
      \label{eq:CGP1}
   \end{equation}
   \begin{equation}
      i\hbar\frac{\pa\psi_{Cs}}{\pa t}=\left(-\frac{\hbar^{2}}{2m_{Cs}}\nabla^{2}+V_{Cs}+g_{Cs}\left|\psi_{Cs}\right|^{2}+g_{RbCs}\left|\psi_{Rb}\right|^{2}-\mu_{Cs}\right)\psi_{Cs},
      \label{eq:CGP2}
    \end{equation}
  \end{widetext}
    where $\psi_{Rb}\left(x,y,z\right)$ and $\psi_{Cs}\left(x,y,z\right)$ are the mean--field wavefunctions for each condensate. Each wavefunction is normalized to its number of atoms, i.e. $\int\left|\psi_{i}\right|^{2}~d x ~d y ~d z=N_{i}$ ($i=\textrm{Rb},~\textrm{Cs}$). The atomic masses and chemical potentials are denoted by $m_{i}$ and $\mu_{i}$. The intraspecies interaction strengths are given by $g_{i}=4\pi\hbar^{2}a_{i}/m_{i}$ and the interspecies interaction strength is $g_\textrm{RbCs}=2\pi\hbar^{2}a_\textrm{RbCs}/M_\textrm{RbCs}$ where $M_\textrm{RbCs}$ is the reduced mass~\cite{PhysRevLett.80.1130}.

    As discussed in~\cite{PhysRevLett.80.1130}, the two components can either overlap (miscible) or phase--separate (immiscible) depending on the relative strength of interactions between the two species. For a homogeneous system, immiscibility requires the interaction strengths to satisfy $g_{12}^{2}>g_{11}g_{22}$~\cite{PhysRevLett.77.3276,Pethick2002}.  While phase separation is suppressed in an inhomogeneous systems due to quantum pressure effects ~\cite{PhysRevA.85.043602}, the Rb--Cs system strongly satisfies the immiscibility criteria and lies deep within the immiscible regime.

We obtain the 3D stationary states of the BEC mixture by solving the CGPEs using the method of steepest descent~\cite{PhysRevA.53.2477} which amounts to simultaneously propagating~\eqref{eq:CGP1} and~\eqref{eq:CGP2} in imaginary time.  As the initial trial solution we employ the independent Thomas-Fermi (TF) density profiles for each condensate~\cite{Pethick2002}.  We employ harmonic oscillator units where time, length and energy are expressed in units of $1/\bar{\omega}_{\textrm{Rb}}=10$ ms, $l_{\rm Rb}=\sqrt{\hbar/m_{Rb}\bar{\omega}_{\textrm{Rb}}}\simeq0.54~\mu$m and $\hbar\bar{\omega}_{\textrm{Rb}}$, respectively, where $\bar{\omega_i}=(\omega_{\perp(i)}^2 \omega_{z(i)})^{1/3}$. We typically present 1D density profiles $n_{1D}(z)$, where the density has been column-integrated in both transverse directions.  

An added complexity of the mean-field model of two-species condensates is the occurrence, for certain parameter regimes, of metastable steady state solutions which can be very close in energy to the true ground state.  These solutions arise from different configurations of the two density profiles.  We find that the steady state solution obtained by imaginary time propagation is strongly dependent on the initial state employed, with the Thomas-Fermi initial states we employ consistently leading to the lowest energy solution, i.e. the ground state.  We demonstrate the existence and behaviour of these metastable solutions in Appendix \ref{1D}, and discuss their presence in relation to our overall results in Sec. \ref{conclude}.  All other results in this work relate to the ground state of the system.  

\subsection{Accounting for trap perturbations}

Given the dominance of the harmonic component of the trapping potential, one may on first inspection anticipate that the ground states under the harmonic trapping of Eq.~(\ref{eqn:V}) would closely match the experimental profiles.  Figure~\ref{fig:points} (c) presents the 1D density profiles of this ground state solution.  As would be expected, the solutions maintain axial symmetry about $z=0$.  For all three cases, we observe an axial structure where the Cs cloud resides at the trap centre with the Rb cloud split either side of it.  The only effect of changing the atom numbers is that relative amplitude of the condensates change.  Our results agree qualitatively with the experimental observations only for Region III, but not those obtained in Regions I and II.  In Region I, the experimental profiles have the reverse structure to our numerical solutions i.e. a central Cs condensate surrounded by Rb, whereas the experimental images for Region II are asymmetric in $z$. The preference in our numerical results for Cs to be centrally positioned is consistent with previous theoretical studies where the component with higher atomic mass resides centrally~\cite{PhysRevLett.80.1130,PhysRevA.62.053601,PhysRevE.65.066201,PhysRevA.70.063606}. 

This dynamical system has a substantial total parameter space; even with a restriction to cylindrical symmetry one is left with eight dimensionless parameters that can in principle be independently varied. Many of the parameters will, in practice, be fixed in any given experimental configuration. Hence, for example, in the experimental configuration described in~\cite{Durham}, it is not possible for the distributions of Rb and Cs to be simply exchanged by changing the particle numbers (the most easily accessible handle to change the system's location in parameter space).  This means, for example that the disagreement of Figs.~\ref{fig:points} (b)(i) and (c)(i) is unlikely to be due to incorrect atom counting.

It is the subject of this paper to study if and how weak anharmonic (spatial) perturbations in the trapping potential may modify the ground state density structures of the system.  We consider the most simple form of perturbation, a linearly varying perturbation.   We apply this perturbation to one species only (Rb) such that the potential experienced by the Rb atoms becomes,
\begin{equation}
V'_\textrm{Rb}(x,y,z)=V_\textrm{Rb}(x,y,z)+\alpha_{z}z+\alpha_{x}x
\label{eqn:Vpert}
\end{equation}
 where $\alpha_{z}$ and $\alpha_{x}$ are the gradients in the axial and {\it one} transverse direction respectively.  The main effect of the linear potential is to shift the trap minimum of $V_\textrm{Rb}$ such that the trap minima for both species no longer coincide, but rather become offset by the distance,
    \begin{equation}
      \delta_{z}=\frac{\alpha_{z}}{m_\textrm{Rb}\omega_{z(\textrm{Rb})}^{2}} \textrm{ and } \delta_{x}=\frac{\alpha_{x}}{m_\textrm{Rb}\omega_{x(\textrm{Rb})}^{2}}.
    \end{equation}ron,D.J. / Ch
We will parametrize the trap perturbations via these distance offsets rather than the linear potential gradients $\alpha_x$ and $\alpha_z$.  We have verified through numerical simulations that we obtain the same results if the harmonic trap centres are instead offset in space, without the addition of linear potentials.

We show that the inclusion of appropriate linear trap perturbations enables us to obtain density structures whose structures match the experimental observations, as shown in Figure \ref{fig:points} (d).

\section{Results: Role of Linear Trap Perturbations \label{offset}}

\begin{figure}
          \subfigure{\includegraphics[width=0.48\textwidth,clip]{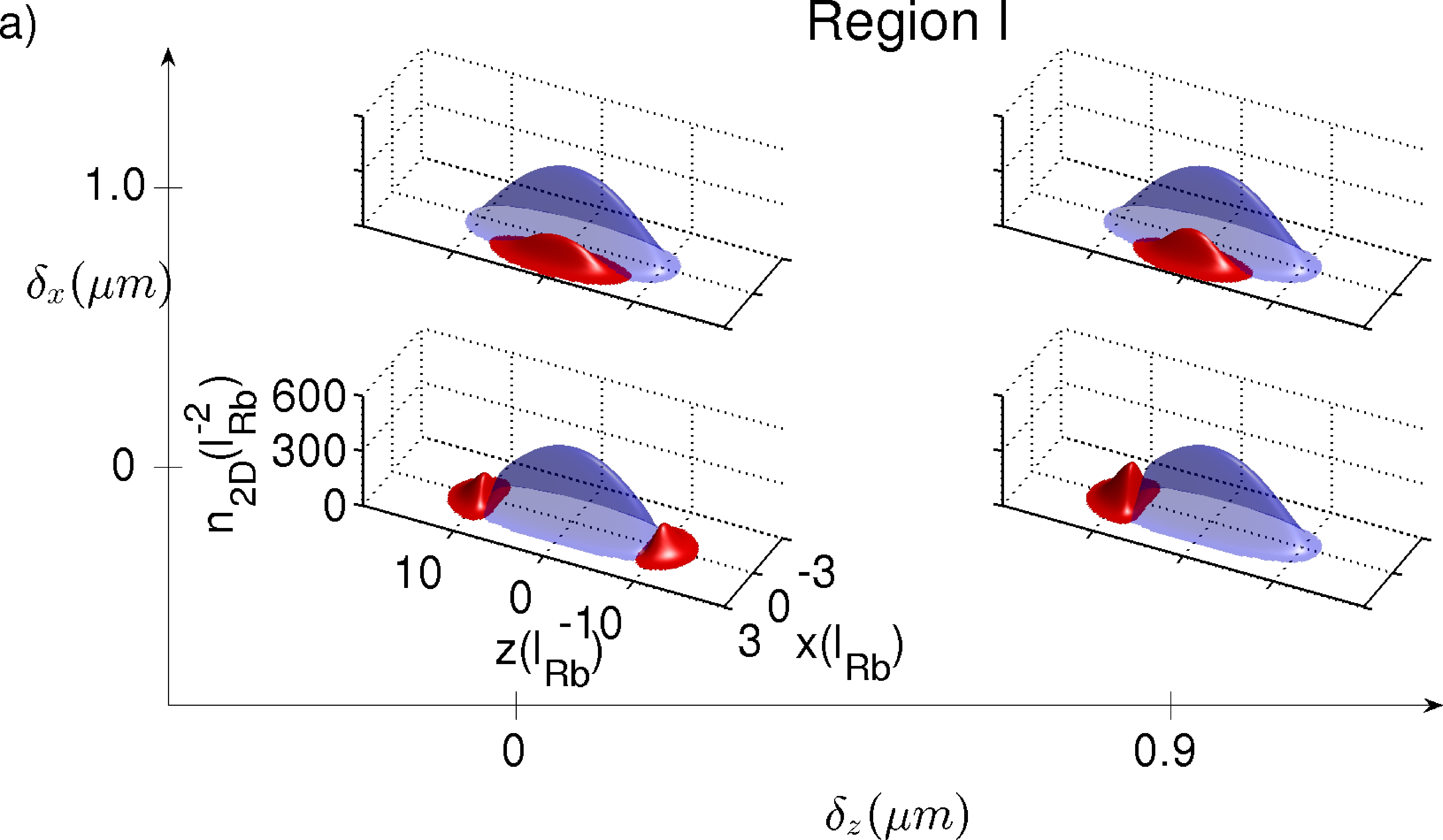}}
          \subfigure{\includegraphics[width=0.48\textwidth,clip]{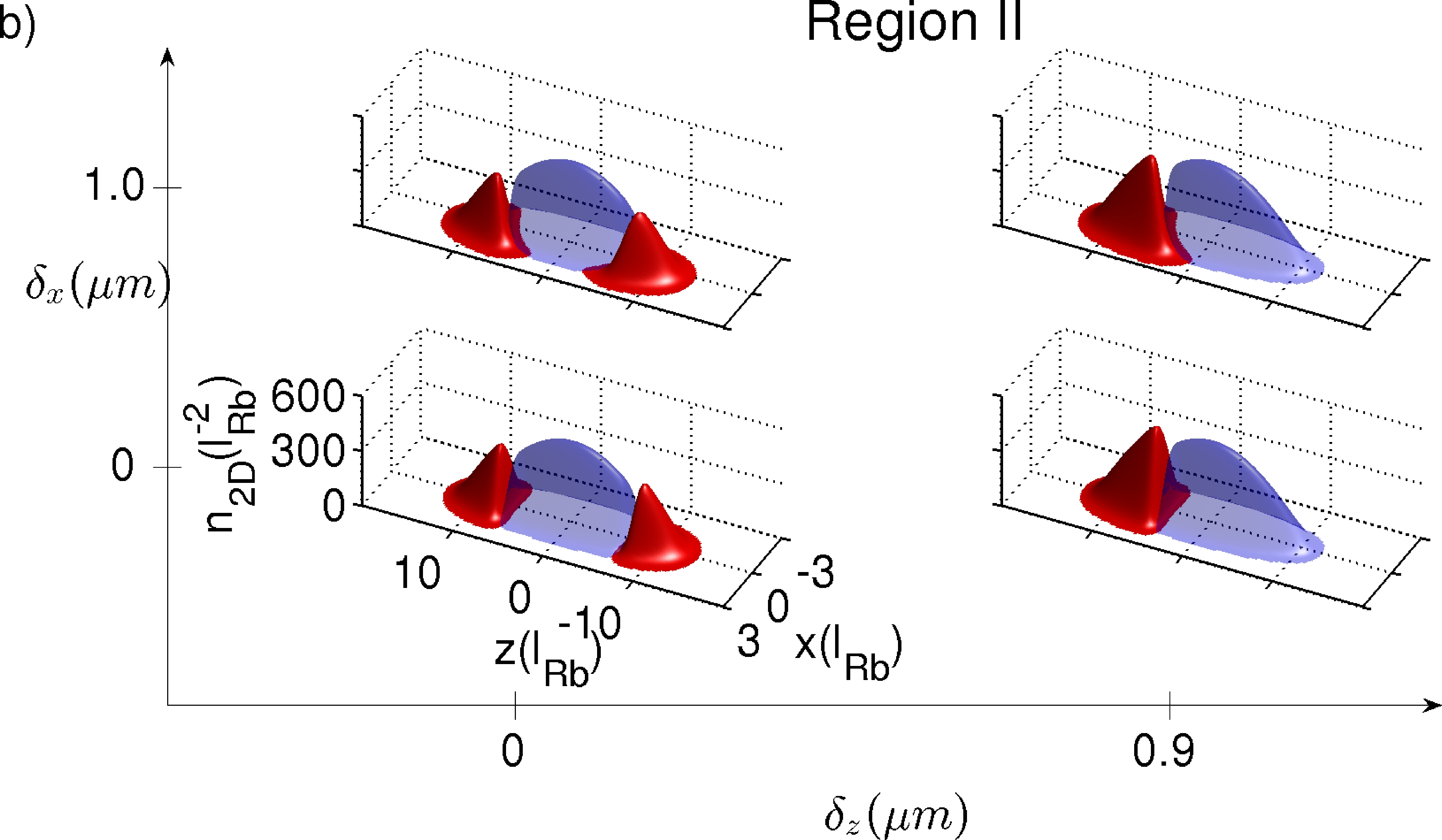}}
          \subfigure{\includegraphics[width=0.48\textwidth,clip]{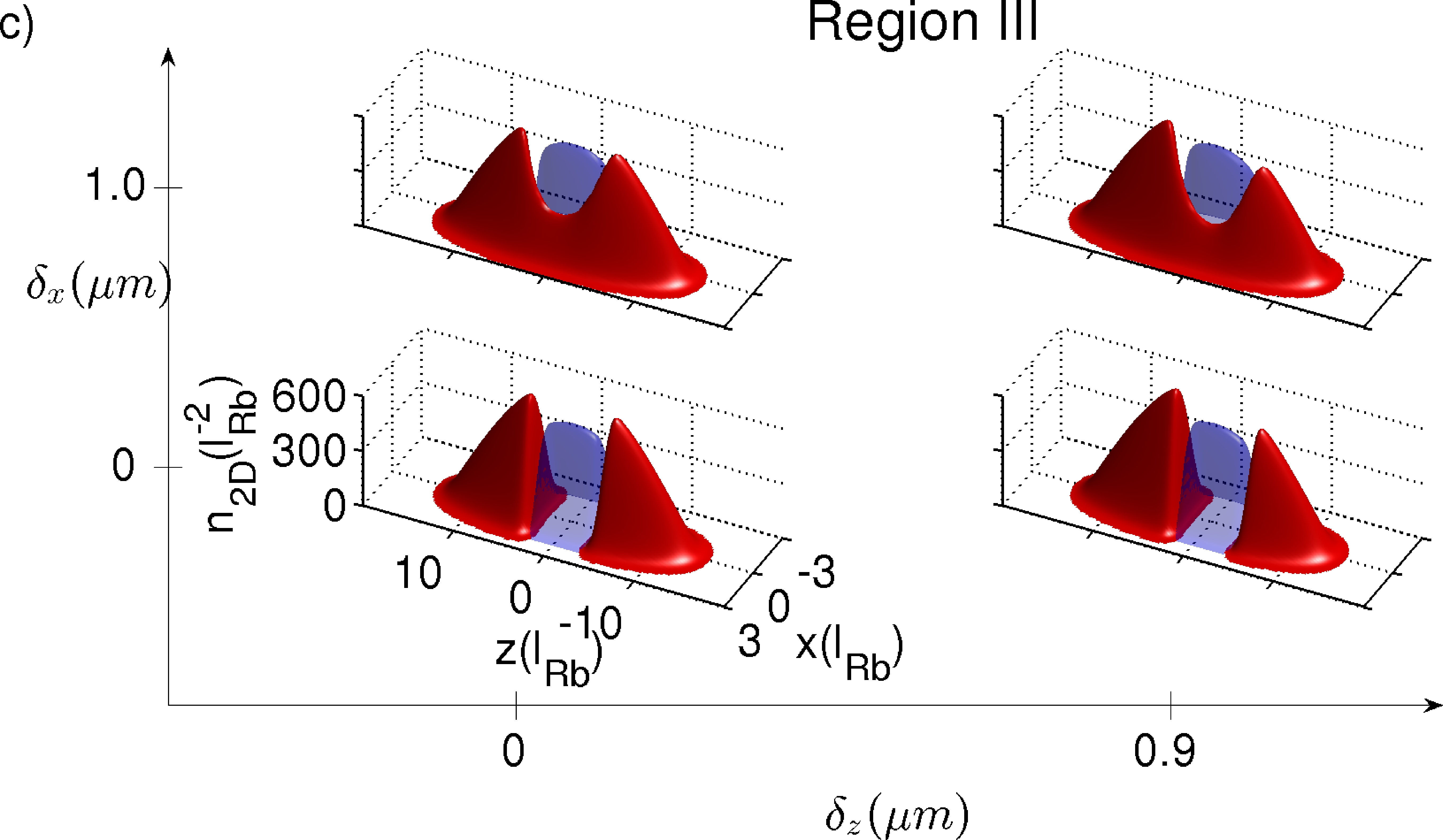}}
          \caption{(Color online) Integrated 2D density profiles of the ground state as a function of the axial (horizontal) and transverse (vertical) offset from top to bottom (a) $N_{\textrm{Rb}}=840$ and $N_{\textrm{Cs}}=8570$, (b) $N_{\textrm{Rb}}=3680$ and $N_{\textrm{Cs}}=8510$, and (c) $N_{\textrm{Rb}}=15100$ and $N_{\textrm{Cs}}=6470$. Red (black) --- Rb; Blue (grey) --- Cs. \label{fig:842}}
        \end{figure}

For the three sets of atom numbers introduced in Sec.~\ref{theory}, we now describe how the ground state solutions are modified by the weak linear trap perturbations to the harmonic trapping potential, according to Eq.~(\ref{eqn:Vpert}). A summary of these results is shown in Fig.~\ref{fig:842} for offsets of  $\delta_{x}=0$ and $1.0~\mu$m, and $\delta_{z}=0$ and $0.9~\mu$m (all combinations thereof). The resulting ground state solutions are found to depend rather sensitively on these displacements and the values chosen have been found to provide the best qualitative agreement with the experimental results, while remaining well within the experimental bounds for the trap perturbations detailed in Sec.~\ref{theory}.  This is the result of a wider analysis of the parameter space of $\delta_x$ and $\delta_z$, where key qualitative effects will be described in the text.  For clarity of how the clouds are distributed in space, we present these results as 2D density profiles in the $x$-$z$ plane, where the density has been integrated in the $y$ direction. The 2D profiles presented here for $\delta_x=\delta_y=0$ correspond to the 1D density profiles presented in Fig.~\ref{fig:points} (c).  

Before discussing the individual behaviours for each regime, there are some general comments to make.  Firstly, for $\delta_x \neq 0$ and $\delta_z \neq 0$, the symmetry in the axial and transverse direction, respectively, become broken thus allowing, in principle, for asymmetric density distributions.  For the fully symmetric potential $\delta_x=\delta_z=0$ the ground state features the Cs cloud sitting centrally, surrounded in $z$ by Rb clouds.  As $\delta_z$ is increased beyond some critical value, one expects the structure to change to a fully asymmetric structure where the Cs and Rb clouds sit side-by-side in the axial direction.  Similarly, as $\delta_x$ is increased past some critical value, one expects the ground state to favour the Rb and Cs clouds sitting side-by-side in the $x$-direction.  We next discuss the specific results for each region in turn.

\subsection{Region I} 
As the transverse offset $\delta_x$ is initially increased from zero, the density structure maintains the same general form (Cs surrounded by two Rb clouds) but with the density profiles becoming skewed transversely (not shown).  For transverse offsets $\delta_{x}\gappeq 0.3~\mu$m, the ground state density suddenly shifts to a transversely side-by-side structure, such as that shown for $\delta_x=1~\mu$m, $\delta_z=0$ in Fig.~\ref{fig:842}. 

On the other hand, as the axial offset $\delta_z$ is increased, the initial structure initially remains but all three clouds become slightly skewed in the $z$ direction.  However, for $\delta_{z}\gappeq 0.4~\mu$m the structure suddenly shifts to being axially side-by side, such as that shown for $\delta_x=0$, $\delta_z=0.9~\mu$m in Fig.~\ref{fig:842}.  In the presence of transverse and axial offsets larger than these critical values, e.g. $\delta_x=1~\mu$m and $\delta_z=0.9\mu$m, the ground state features the Cs and Rb in the transversely side-by-side structure, with both slightly skewed in $z$.  Importantly, the corresponding 1D density profile shown in Fig.~\ref{fig:points} (d) now bears the same qualitative structure as the experiment results with the Rb sitting centrally and inducing a weak depression in the integrated Cs density. We rule here that it appears to be crucial that $\delta_{z}<\delta_{x}$ for such features to form and that the ground state for $\delta_{z}=\delta_{x}=1\mu m$ is actually axially asymmetric.

\subsection{Region II} 
For this case, a transverse offset of $\delta_x=1~\mu$m does {\it not} give rise to a transverse side-by-side ground state; the system remains in a three peak configuration with Rb clouds either side of the central Cs cloud.  A larger transverse perturbation $\delta_{x}\simeq 1.5~\mu$m is required to induce a shift to a transverse side-by-side state.  In the axial direction, an offset of  $\delta_{z}\simeq 0.5~\mu$m is sufficient to give rise to axially side-by-side density profiles similar to Region I. The combination of offsets in both directions leaves the condensates in an axially asymmetric configuration as with no transverse displacement. The ground state density profiles for this set of atom numbers is more sensitive to the addition of axial linear potentials to the harmonic trap in comparison to the transverse ones. The corresponding 1D density profile shown in Fig.~\ref{fig:points} (d) now features the side-by-side structure observed in the corresponding experimental image.   

\subsection{Region III} 
Here, the ground state always remains in a three--peak configuration for small offsets.  Under the addition of transverse offset the Rb cloud is no longer split into two un-connected sections but rather joins on one side of the Cs cloud.  Under the addition of the axial offset the amplitudes of the Rb peaks becomes asymmetric. A blend of both of these affects is seen when a combination of the small offsets due to the additional linear potentials in both the transverse and axial directions is used. Side--by--side density profiles can be obtained in the transverse and axial directions for much larger offsets (which are beyond bounds for the experiment in particular) of $\delta_{x}\simeq 2.0~\mu$m and $\delta_{z}\simeq 4.0~\mu$m, respectively.  The corresponding 1D density profile shown in Fig.~\ref{fig:points} (d) has a similar structure to the original, fully symmetric result but with an increased central density for Rb and a skewed axial profile.  As such, it maintains the same qualitative structure to the corresponding experiment image.

\subsection{Overall behaviour}
To summarise the above results, the combination of offsets $\delta_x=1$ and $\delta_z=0.9~\mu$m leads to the optimum comparison to the experimental results, in which the mean-field ground state recovers the three density structures observed experimentally.   In Region I, the central density dip in the Cs profile is more pronounced in the experimental observations, e.g. Fig.~\ref{fig:points} (b)(i), than in the theoretical results.  An inherent feature of solving the coupled Gross-Pitaevskii equations for immiscible two-species BEC is a sensitivity to the initial trial wavefunction.  This is further discussed within Appendix A and complimentary results are presented in Ref.~\cite{PhysRevA.85.023613}.  All of our results presented so far have been based on TF initial trial wavefunctions, as described in Sec.~\ref{theory}.  By their nature, the TF profiles tend to be broadly distributed in space, which favours a broader overall density distribution in the final static solution. We find that employing an initial distribution for the Rb cloud which is tightly localized at the origin yields static solutions which retains the same features as before but with a more prominent density dip in the adjacent Cs cloud, in closer agreement with the experimental profiles for Region I.

We have also looked at introducing trap offsets in our initial conditions only (without permanent trap shifts) whereby the TF initial conditions for each species are initially offset along the $z$ and/or $x$--axis. Similarly to the use of the linear potential, this initial offset could be tailored to reproduce the experimental results to a comparable degree of accuracy.

We have additionally simulated the expansion of the static solutions following the sudden removal of all trapping potentials.  This expansion is performed in the experiments prior to imaging.  Our analysis showed that expansion does not affect the structures formed. The overall phase separation features appeared to be captured very well under the assumption made here that the profiles observed in the experiments are the true equilibrium profiles and that these profiles are dominated by their respective condensate component, with thermal clouds simply modifying these profiles by the addition of characteristic thermal tails.

\section{Discussion\label{conclude}}

We investigated the $^{87}$Rb--$^{133}$Cs ground state density profiles corresponding to the parameters of a recent experiment [D.~J. McCarron {\em et al.}, Phys. Rev. A {\bf 84}, 011603 (2011)].  This was conducted within the simplest possible zero--temperature mean--field theory consisting of two coupled Gross--Pitaevskii equations. Density profiles obtained in perfectly symmetric traps were found not to match the experimental results. Analysing the experiment more carefully, we proceeded to add weak perturbations to the harmonic trap (in the form of linear potentials) in the axial and one transverse direction accounting for anticipated experimental offsets (of around 1 $\mu$m) in the trap centres for the two species.  Even weak trap perturbations can give rise to dramatically different density profiles.  Importantly, this allows us to obtain the observed asymmetric experimental profiles. In particular we found that the axial shift needed to be slightly smaller than the transverse one for such features to be numerically obtained. By tailoring the size of the perturbations, we found our simulations to qualitatively match structural regimes seen experimentally when focusing on condensate phase separation features (and overlooking the experimental existence of thermal tails which is not accounted for in our model).

The analysis presented in this work was based on equilibrium density profiles. While we demonstrated good overall agreement with the experimentally-reported profiles, we also found that a change in the initial conditions of the simulations, e.g.\ one of the components being more tightly localised in the centre, could affect the final equilibrated profiles, as numerous metastable states (of comparable, but not identical, energies) exist for each configuration. Such a situation could for example arise in the early stages of coupled growth under some parameter regimes. In the experiments, as the two species were sympathetically cooled, the initial number of condensate atoms within each species (or the sequence by which growth proceeded) was not accurately known. Moreover, the density profiles were typically measured after a variable hold time, without necessarily guaranteeing that the structures observed were indeed true equilibrium states (as opposed to some long-lived metastable steady-states), for which a detailed analysis of growth dynamics would have been required.
Preliminary investigation of coupled Gross--Pitaevskii equations with phenomenological damping undertaken by us indeed revealed different features during growth, depending on both initial conditions and growth parameters; more importantly, however, such simulations showed that after sufficient evolution time, the condensate in one or the other species disappeared, a feature which is in qualitative agreement with the experiments, which detected only one condensate (of either species) in some measurements.
The study of coupled two-component condensate growth is an interesting topic that will be studied in more detail in future work.
Similar non-equilibrium conclusions have been reached by another group~\cite{post} using such equations additionally modified by the presence of stochastic noise mimicking thermal fluctuations, which additionally allows for the appearance of spontaneous structures during growth.

  \begin{acknowledgements}
    We thank the UK EPSRC (grants EP/G056781/1, GR/S78339/01 and EP/H003363/1) for support.
We also acknowledge discussions with I.-Kang. Liu and Shih-Chuan Gou during the writing-up phase of this work.
 \end{acknowledgements}

  \appendix
    \section{Sensitivity to Initial Conditions\label{1D}}

      In this appendix, we illustrate how the trapped immiscible two-species condensates can possess a family of stationary states of similar energy that can be accessed through varying the initial conditions for imaginary-time convergence of the CGPEs. For simplicity we focus on a generic 1D two-species system described by 1D CGPEs, in which the transverse wavefunctions are assumed to be the ground Gaussian harmonic oscillator states of width $l_{\perp(i)}=\sqrt{\hbar/m \omega_{\perp(i)}}$ ($i=1,2$).  The 1D interactions, denoted by $U$, are given by $U_{ii}=g_{ii}/2\pi l^{2}_{\perp(i)}$, $U_{12}=g_{12}/\pi\left(l^{2}_{\perp (i)}+l^{2}_{\perp (i)}\right)$ and $\mu_\textrm{1D (i)}=\mu_{i}-\hbar\omega_{\perp (i)}$. Similarly to~\cite{Tripp}, we take $N=N_{1}=N_{2}$, $\omega_{1}=\omega_{2}$, $m_{1}=m_{2}$ and $U_{22}=1.01U_{11}$, $U_{12}=1.52U_{11}$.  The two traps are co-centred in space. 

       \begin{figure}
        \includegraphics[width=0.48\textwidth,clip]{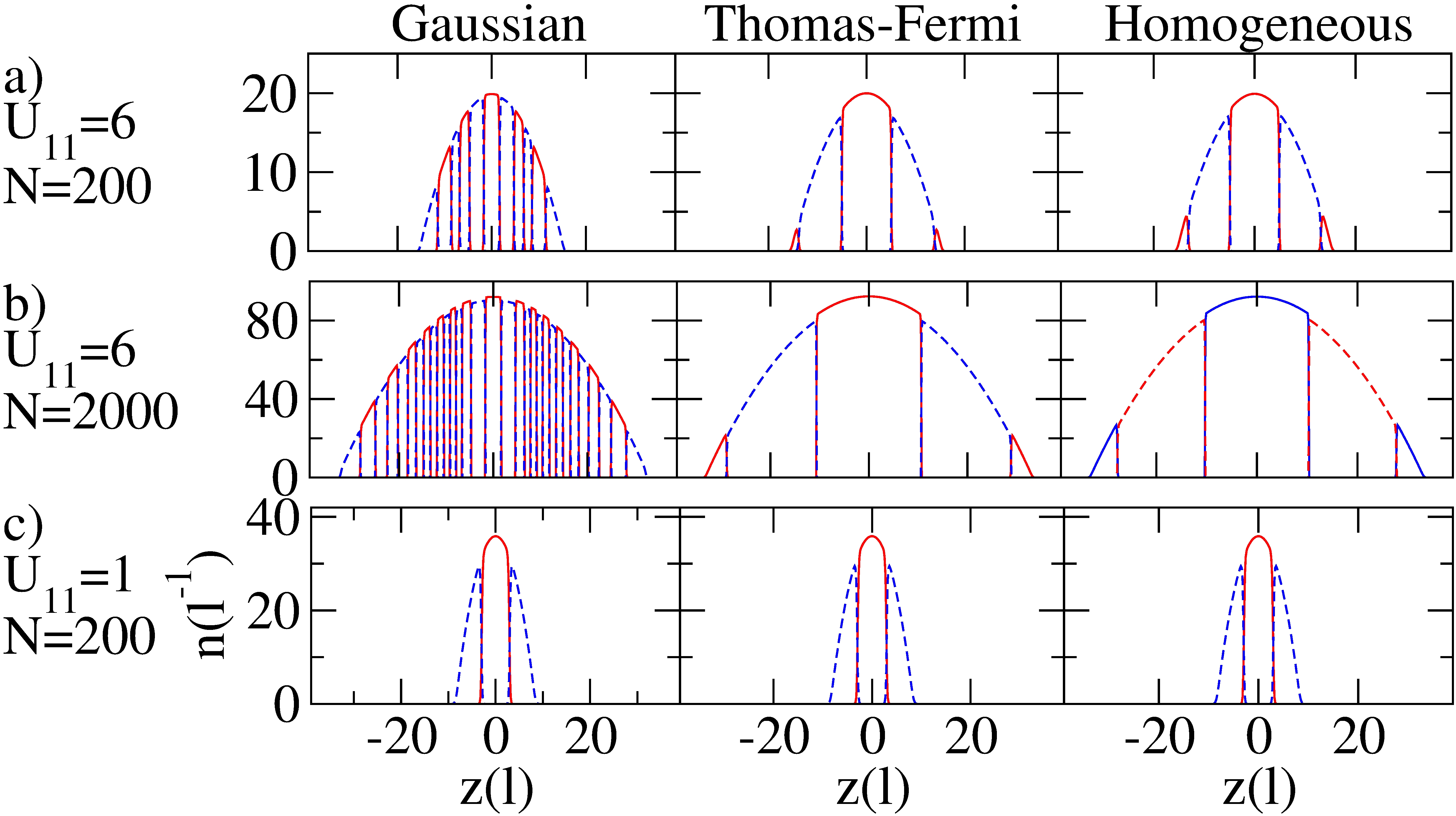}
        \caption{(color online) 1D density profiles of stationary states of the 1D CGPEs with $N=N_{1}=N_{2}$, $\omega_{1}=\omega_{2}$, $m_{1}=m_{2}$ and $U_{22}=1.01U_{11}$, $U_{12}=1.52U_{11}$. Columns correspond to Gaussian, TF and homogeneous initial conditions for imaginary time propagation. (a) $U_{11}=6$, $N=200$. (b) $U_{11}=6$, $N=2000$. (c) $U_{11}=1$, $N=200$.
Solid blue curve -- species 1; Dashed red curve -- species 2.\label{fig:1D}}
      \end{figure}
      \begin{figure}
        \includegraphics[width=0.48\textwidth,clip]{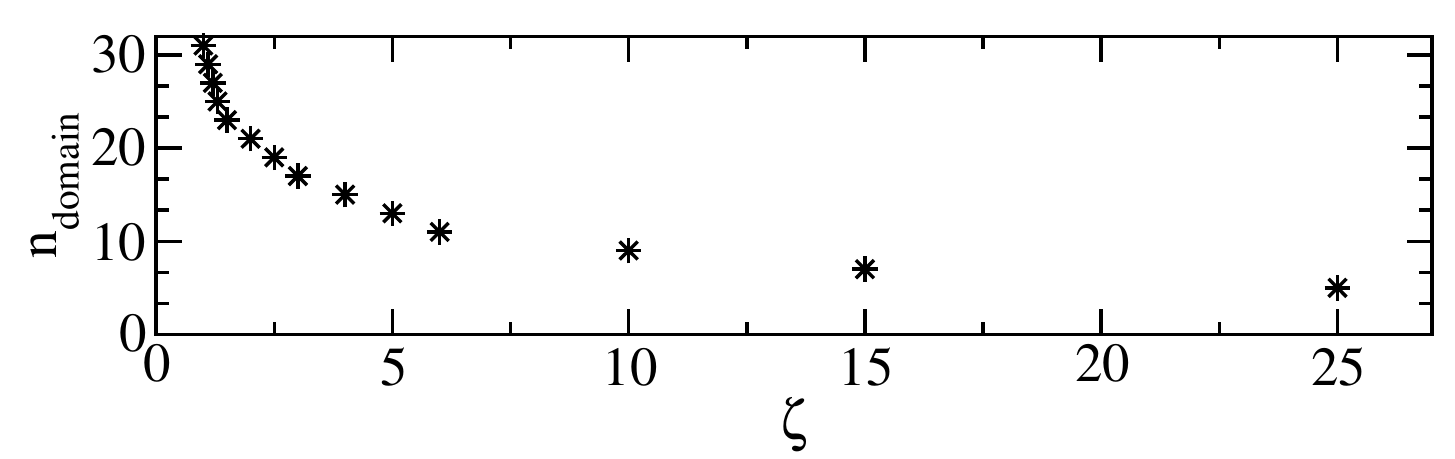}
        \caption{Number of density domains $n_{domain}$ in the obtained stationary state as a function of the width of the Gaussian initial condition $\zeta$ (with $\zeta=1$ corresponding to the Gaussian ground harmonic oscillator state), for the system parameters of Fig.~\ref{fig:1D}(b).\label{fig:1DGauss}}
      \end{figure}

 We firstly consider initial conditions which are the (i) Gaussian ground harmonic oscillator state, (ii) Thomas-Fermi (TF) solution, and (iii) homogeneous (uniform density) state, for each species.  The converged stationary states, following imaginary time propagation, are shown in Fig. \ref{fig:1D}.  Sensitivity to the initial condition is evident.  The TF and homogeneous initial conditions favour a stationary state with a few density domains, while the Gaussian-derived stationary state contains many more domains.  This effect increases with the non-linearity, i.e. increasing atom number and/or interaction strength.  For weak non-linearity (low atom number, weak interactions) this effect becomes washed out and all initial conditions lead to the ground stationary state.  Where sensitivity to initial condition does  occur, the TF-derived state is the ground (lowest energy) state while the Gaussian-derived solution has the greatest energy.   The difference in energies is small, typically less than $10$\%.  

The formation of states with an increased number of domains is attributed to a modulational instability of the condensates during imaginary-time propagation.   This instability is highly sensitive to the spatial extent of the initial conditions.  To illustrate this we consider a Gaussian initial condition of width $\zeta \ell$, where $\ell$ denotes the axial harmonic oscillator length.  We introduce the parameter $n_{domain}$ giving the number of density domains in the stationary solution.  In Fig.~\ref{fig:1DGauss}(a) we plot $n_{domain}$ as a function of the Gaussian width $\zeta$ (for the system parameters of Fig.~\ref{fig:1D}(b)).  For a wide Gaussian initial state ($\zeta\geq25$), we obtain a stationary state with 3 domains - one species in the trap centre surrounded by the other species.   As $\zeta$ is reduced, $n_{domain}$ increases exponentially, appearing to diverge as $\zeta \rightarrow 0$.

In 3D the stationary states obtained are also sensitive to initial conditions.  Although the sensitivity is less than in 1D, we can use the initial conditions as a handle to match the experimental density profiles.  For example, recall Fig.~\ref{fig:points} (c)(i), a stationary (ground) state (derived using TF initial conditions) in which the Rb sits either side of the central Cs cloud.  If we instead use a very narrow Gaussian profile for the Rb initial condition (retaining the TF profile for Cs) we can numerically converge to an excited stationary state which features the Rb sitting at the trap centre and a small density dip in the ambient Cs cloud, in qualitative agreement with the corresponding experimental profile (without the inclusion of trap perturbations).  This initial condition could correspond to the physical situation where a TF-like Cs condensate is pre-formed after which the Rb atoms begin to condense in a small narrow region at the center of the traps, creating an initially very localised condensate at the trap center.


\begin{thebibliography}{10}%
\makeatletter
\providecommand \@ifxundefined [1]{%
 \ifx #1\undefined \expandafter \@firstoftwo
 \else \expandafter \@secondoftwo
\fi
}%
\providecommand \@ifnum [1]{%
 \ifnum #1\expandafter \@firstoftwo
 \else \expandafter \@secondoftwo
\fi
}%
\providecommand \enquote [1]{``#1''}%
\providecommand \bibnamefont  [1]{#1}%
\providecommand \bibfnamefont [1]{#1}%
\providecommand \citenamefont [1]{#1}%
\providecommand\href[0]{\@sanitize\@href}%
\providecommand\@href[1]{\endgroup\@@startlink{#1}\endgroup\@@href}%
\providecommand\@@href[1]{#1\@@endlink}%
\providecommand \@sanitize [0]{\begingroup\catcode`\&12\catcode`\#12\relax}%
\@ifxundefined \pdfoutput {\@firstoftwo}{%
 \@ifnum{\z@=\pdfoutput}{\@firstoftwo}{\@secondoftwo}%
}{%
 \providecommand\@@startlink[1]{\leavevmode}%
 \providecommand\@@endlink[0]{}%
}{%
 \providecommand\@@startlink[1]{%
  \leavevmode
  \pdfstartlink
   attr{/Border[0 0 1 ]/H/I/C[0 1 1]}%
   user{/Subtype/Link/A<</Type/Action/S/URI/URI(#1)>>}%
  \relax
 }%
 \providecommand\@@endlink[0]{\pdfendlink}%
}%
\providecommand \url  [0]{\begingroup\@sanitize \@url }%
\providecommand \@url [1]{\endgroup\@href {#1}{\urlprefix}}%
\providecommand \urlprefix [0]{URL }%
\providecommand \Eprint[0]{\href }%
\@ifxundefined \urlstyle {%
  \providecommand \doi [1]{doi:\discretionary{}{}{}#1}%
}{%
  \providecommand \doi [0]{doi:\discretionary{}{}{}\begingroup
  \urlstyle{rm}\Url }%
}%
\providecommand \doibase [0]{http://dx.doi.org/}%
\providecommand \Doi[1]{\href{\doibase#1}}%
\providecommand \bibAnnote [3]{%
  \BibitemShut{#1}%
  \begin{quotation}\noindent
    \textsc{Key:}\ #2\\\textsc{Annotation:}\ #3%
  \end{quotation}%
}%
\providecommand \bibAnnoteFile [2]{%
  \IfFileExists{#2}{\bibAnnote {#1} {#2} {\input{#2}}}{}%
}%
\providecommand \typeout [0]{\immediate \write \m@ne }%
\providecommand \selectlanguage [0]{\@gobble}%
\providecommand \bibinfo [0]{\@secondoftwo}%
\providecommand \bibfield [0]{\@secondoftwo}%
\providecommand \translation [1]{[#1]}%
\providecommand \BibitemOpen[0]{}%
\providecommand \bibitemStop [0]{}%
\providecommand \bibitemNoStop [0]{.\EOS\space}%
\providecommand \EOS [0]{\spacefactor3000\relax}%
\providecommand \BibitemShut [1]{\csname bibitem#1\endcsname}%
%</preamble>
\bibitem{PhysRevLett.78.586}%
  \BibitemOpen
  \bibfield{author}{%
  \bibinfo {author} {\bibfnamefont{C.~J.}\ \bibnamefont{Myatt}}, \bibinfo
  {author} {\bibfnamefont{E.~A.}\ \bibnamefont{Burt}}, \bibinfo {author}
  {\bibfnamefont{R.~W.}\ \bibnamefont{Ghrist}}, \bibinfo {author}
  {\bibfnamefont{E.~A.}\ \bibnamefont{Cornell}},\ and\ \bibinfo {author}
  {\bibfnamefont{C.~E.}\ \bibnamefont{Wieman}},\ }%
  \bibfield{journal}{%
  \Doi{10.1103/PhysRevLett.78.586}{\bibinfo {journal} {Phys. Rev. Lett.}}\ }%
  \textbf{\bibinfo {volume} {78}},\ \bibinfo {pages} {586} (\bibinfo {month}
  {Jan}\ \bibinfo {year} {1997})%
  \bibAnnoteFile{NoStop}{PhysRevLett.78.586}%
\bibitem{PhysRevLett.89.053202}%
  \BibitemOpen
  \bibfield{author}{%
  \bibinfo {author} {\bibfnamefont{G.}~\bibnamefont{Ferrari}}, \bibinfo
  {author} {\bibfnamefont{M.}~\bibnamefont{Inguscio}}, \bibinfo {author}
  {\bibfnamefont{W.}~\bibnamefont{Jastrzebski}}, \bibinfo {author}
  {\bibfnamefont{G.}~\bibnamefont{Modugno}}, \bibinfo {author}
  {\bibfnamefont{G.}~\bibnamefont{Roati}},\ and\ \bibinfo {author}
  {\bibfnamefont{A.}~\bibnamefont{Simoni}},\ }%
  \bibfield{journal}{%
  \Doi{10.1103/PhysRevLett.89.053202}{\bibinfo {journal} {Phys. Rev. Lett.}}\
  }%
  \textbf{\bibinfo {volume} {89}},\ \bibinfo {pages} {053202} (\bibinfo {month}
  {Jul}\ \bibinfo {year} {2002})%
  \bibAnnoteFile{NoStop}{PhysRevLett.89.053202}%
\bibitem{PhysRevLett.89.190404}%
  \BibitemOpen
  \bibfield{author}{%
  \bibinfo {author} {\bibfnamefont{G.}~\bibnamefont{Modugno}}, \bibinfo
  {author} {\bibfnamefont{M.}~\bibnamefont{Modugno}}, \bibinfo {author}
  {\bibfnamefont{F.}~\bibnamefont{Riboli}}, \bibinfo {author}
  {\bibfnamefont{G.}~\bibnamefont{Roati}},\ and\ \bibinfo {author}
  {\bibfnamefont{M.}~\bibnamefont{Inguscio}},\ }%
  \bibfield{journal}{%
  \Doi{10.1103/PhysRevLett.89.190404}{\bibinfo {journal} {Phys. Rev. Lett.}}\
  }%
  \textbf{\bibinfo {volume} {89}},\ \bibinfo {pages} {190404} (\bibinfo {month}
  {Oct}\ \bibinfo {year} {2002})%
  \bibAnnoteFile{NoStop}{PhysRevLett.89.190404}%
\bibitem{PhysRevLett.100.210402}%
  \BibitemOpen
  \bibfield{author}{%
  \bibinfo {author} {\bibfnamefont{G.}~\bibnamefont{Thalhammer}}, \bibinfo
  {author} {\bibfnamefont{G.}~\bibnamefont{Barontini}}, \bibinfo {author}
  {\bibfnamefont{L.}~\bibnamefont{De~Sarlo}}, \bibinfo {author}
  {\bibfnamefont{J.}~\bibnamefont{Catani}}, \bibinfo {author}
  {\bibfnamefont{F.}~\bibnamefont{Minardi}},\ and\ \bibinfo {author}
  {\bibfnamefont{M.}~\bibnamefont{Inguscio}},\ }%
  \bibfield{journal}{%
  \Doi{10.1103/PhysRevLett.100.210402}{\bibinfo {journal} {Phys. Rev. Lett.}}\
  }%
  \textbf{\bibinfo {volume} {100}},\ \bibinfo {pages} {210402} (\bibinfo
  {month} {May}\ \bibinfo {year} {2008})%
  \bibAnnoteFile{NoStop}{PhysRevLett.100.210402}%
\bibitem{Durham}%
  \BibitemOpen
  \bibfield{author}{%
  \bibinfo {author} {\bibfnamefont{D.~J.}\ \bibnamefont{McCarron}}, \bibinfo
  {author} {\bibfnamefont{H.~W.}\ \bibnamefont{Cho}}, \bibinfo {author}
  {\bibfnamefont{D.~L.}\ \bibnamefont{Jenkin}}, \bibinfo {author}
  {\bibfnamefont{M.~P.}\ \bibnamefont{K\"oppinger}},\ and\ \bibinfo {author}
  {\bibfnamefont{S.~L.}\ \bibnamefont{Cornish}},\ }%
  \bibfield{journal}{%
  \Doi{10.1103/PhysRevA.84.011603}{\bibinfo {journal} {Phys. Rev. A}}\ }%
  \textbf{\bibinfo {volume} {84}},\ \bibinfo {pages} {011603} (\bibinfo {month}
  {Jul}\ \bibinfo {year} {2011})%
  \bibAnnoteFile{NoStop}{Durham}%
\bibitem{PhysRevLett.101.040402}%
  \BibitemOpen
  \bibfield{author}{%
  \bibinfo {author} {\bibfnamefont{S.~B.}\ \bibnamefont{Papp}}, \bibinfo
  {author} {\bibfnamefont{J.~M.}\ \bibnamefont{Pino}},\ and\ \bibinfo {author}
  {\bibfnamefont{C.~E.}\ \bibnamefont{Wieman}},\ }%
  \bibfield{journal}{%
  \Doi{10.1103/PhysRevLett.101.040402}{\bibinfo {journal} {Phys. Rev. Lett.}}\
  }%
  \textbf{\bibinfo {volume} {101}},\ \bibinfo {pages} {040402} (\bibinfo
  {month} {Jul}\ \bibinfo {year} {2008})%
  \bibAnnoteFile{NoStop}{PhysRevLett.101.040402}%
\bibitem{PhysRevLett.81.1539}%
  \BibitemOpen
  \bibfield{author}{%
  \bibinfo {author} {\bibfnamefont{D.~S.}\ \bibnamefont{Hall}}, \bibinfo
  {author} {\bibfnamefont{M.~R.}\ \bibnamefont{Matthews}}, \bibinfo {author}
  {\bibfnamefont{J.~R.}\ \bibnamefont{Ensher}}, \bibinfo {author}
  {\bibfnamefont{C.~E.}\ \bibnamefont{Wieman}},\ and\ \bibinfo {author}
  {\bibfnamefont{E.~A.}\ \bibnamefont{Cornell}},\ }%
  \bibfield{journal}{%
  \Doi{10.1103/PhysRevLett.81.1539}{\bibinfo {journal} {Phys. Rev. Lett.}}\ }%
  \textbf{\bibinfo {volume} {81}},\ \bibinfo {pages} {1539} (\bibinfo {month}
  {Aug}\ \bibinfo {year} {1998})%
  \bibAnnoteFile{NoStop}{PhysRevLett.81.1539}%
\bibitem{PhysRevLett.83.2498}%
  \BibitemOpen
  \bibfield{author}{%
  \bibinfo {author} {\bibfnamefont{M.~R.}\ \bibnamefont{Matthews}}, \bibinfo
  {author} {\bibfnamefont{B.~P.}\ \bibnamefont{Anderson}}, \bibinfo {author}
  {\bibfnamefont{P.~C.}\ \bibnamefont{Haljan}}, \bibinfo {author}
  {\bibfnamefont{D.~S.}\ \bibnamefont{Hall}}, \bibinfo {author}
  {\bibfnamefont{C.~E.}\ \bibnamefont{Wieman}},\ and\ \bibinfo {author}
  {\bibfnamefont{E.~A.}\ \bibnamefont{Cornell}},\ }%
  \bibfield{journal}{%
  \Doi{10.1103/PhysRevLett.83.2498}{\bibinfo {journal} {Phys. Rev. Lett.}}\ }%
  \textbf{\bibinfo {volume} {83}},\ \bibinfo {pages} {2498} (\bibinfo {month}
  {Sep}\ \bibinfo {year} {1999})%
  \bibAnnoteFile{NoStop}{PhysRevLett.83.2498}%
\bibitem{PhysRevLett.85.2413}%
  \BibitemOpen
  \bibfield{author}{%
  \bibinfo {author} {\bibfnamefont{P.}~\bibnamefont{Maddaloni}}, \bibinfo
  {author} {\bibfnamefont{M.}~\bibnamefont{Modugno}}, \bibinfo {author}
  {\bibfnamefont{C.}~\bibnamefont{Fort}}, \bibinfo {author}
  {\bibfnamefont{F.}~\bibnamefont{Minardi}},\ and\ \bibinfo {author}
  {\bibfnamefont{M.}~\bibnamefont{Inguscio}},\ }%
  \bibfield{journal}{%
  \Doi{10.1103/PhysRevLett.85.2413}{\bibinfo {journal} {Phys. Rev. Lett.}}\ }%
  \textbf{\bibinfo {volume} {85}},\ \bibinfo {pages} {2413} (\bibinfo {month}
  {Sep}\ \bibinfo {year} {2000})%
  \bibAnnoteFile{NoStop}{PhysRevLett.85.2413}%
\bibitem{PhysRevA.63.051602}%
  \BibitemOpen
  \bibfield{author}{%
  \bibinfo {author} {\bibfnamefont{G.}~\bibnamefont{Delannoy}}, \bibinfo
  {author} {\bibfnamefont{S.~G.}\ \bibnamefont{Murdoch}}, \bibinfo {author}
  {\bibfnamefont{V.}~\bibnamefont{Boyer}}, \bibinfo {author}
  {\bibfnamefont{V.}~\bibnamefont{Josse}}, \bibinfo {author}
  {\bibfnamefont{P.}~\bibnamefont{Bouyer}},\ and\ \bibinfo {author}
  {\bibfnamefont{A.}~\bibnamefont{Aspect}},\ }%
  \bibfield{journal}{%
  \Doi{10.1103/PhysRevA.63.051602}{\bibinfo {journal} {Phys. Rev. A}}\ }%
  \textbf{\bibinfo {volume} {63}},\ \bibinfo {pages} {051602} (\bibinfo {month}
  {Apr}\ \bibinfo {year} {2001})%
  \bibAnnoteFile{NoStop}{PhysRevA.63.051602}%
%\bibitem{PhysRevLett.93.210403}%
%  \BibitemOpen
%  \bibfield{author}{%
%  \bibinfo {author} {\bibfnamefont{V.}~\bibnamefont{Schweikhard}}, \bibinfo
%  {author} {\bibfnamefont{I.}~\bibnamefont{Coddington}}, \bibinfo {author}
%  {\bibfnamefont{P.}~\bibnamefont{Engels}}, \bibinfo {author}
%  {\bibfnamefont{S.}~\bibnamefont{Tung}},\ and\ \bibinfo {author}
%  {\bibfnamefont{E.~A.}\ \bibnamefont{Cornell}},\ }%
%  \bibfield{journal}{%
%  \Doi{10.1103/PhysRevLett.93.210403}{\bibinfo {journal} {Phys. Rev. Lett.}}\
%  }%
%  \textbf{\bibinfo {volume} {93}},\ \bibinfo {pages} {210403} (\bibinfo {month}
%  {Nov}\ \bibinfo {year} {2004})%
%  \bibAnnoteFile{NoStop}{PhysRevLett.93.210403}%
\bibitem{PhysRevLett.99.190402}%
  \BibitemOpen
  \bibfield{author}{%
  \bibinfo {author} {\bibfnamefont{K.~M.}\ \bibnamefont{Mertes}}, \bibinfo
  {author} {\bibfnamefont{J.~W.}\ \bibnamefont{Merrill}}, \bibinfo {author}
  {\bibfnamefont{R.}~\bibnamefont{Carretero-Gonz\'alez}}, \bibinfo {author}
  {\bibfnamefont{D.~J.}\ \bibnamefont{Frantzeskakis}}, \bibinfo {author}
  {\bibfnamefont{P.~G.}\ \bibnamefont{Kevrekidis}},\ and\ \bibinfo {author}
  {\bibfnamefont{D.~S.}\ \bibnamefont{Hall}},\ }%
  \bibfield{journal}{%
  \Doi{10.1103/PhysRevLett.99.190402}{\bibinfo {journal} {Phys. Rev. Lett.}}\
  }%
  \textbf{\bibinfo {volume} {99}},\ \bibinfo {pages} {190402} (\bibinfo {month}
  {Nov}\ \bibinfo {year} {2007})%
  \bibAnnoteFile{NoStop}{PhysRevLett.99.190402}%
\bibitem{PhysRevA.80.023603}%
  \BibitemOpen
  \bibfield{author}{%
  \bibinfo {author} {\bibfnamefont{R.~P.}\ \bibnamefont{Anderson}}, \bibinfo
  {author} {\bibfnamefont{C.}~\bibnamefont{Ticknor}}, \bibinfo {author}
  {\bibfnamefont{A.~I.}\ \bibnamefont{Sidorov}},\ and\ \bibinfo {author}
  {\bibfnamefont{B.~V.}\ \bibnamefont{Hall}},\ }%
  \bibfield{journal}{%
  \Doi{10.1103/PhysRevA.80.023603}{\bibinfo {journal} {Phys. Rev. A}}\ }%
  \textbf{\bibinfo {volume} {80}},\ \bibinfo {pages} {023603} (\bibinfo {month}
  {Aug}\ \bibinfo {year} {2009})%
  \bibAnnoteFile{NoStop}{PhysRevA.80.023603}%
\bibitem{PhysRevA.82.033609}%
  \BibitemOpen
  \bibfield{author}{%
  \bibinfo {author} {\bibfnamefont{S.}~\bibnamefont{Tojo}}, \bibinfo {author}
  {\bibfnamefont{Y.}~\bibnamefont{Taguchi}}, \bibinfo {author}
  {\bibfnamefont{Y.}~\bibnamefont{Masuyama}}, \bibinfo {author}
  {\bibfnamefont{T.}~\bibnamefont{Hayashi}}, \bibinfo {author}
  {\bibfnamefont{H.}~\bibnamefont{Saito}},\ and\ \bibinfo {author}
  {\bibfnamefont{T.}~\bibnamefont{Hirano}},\ }%
  \bibfield{journal}{%
  \Doi{10.1103/PhysRevA.82.033609}{\bibinfo {journal} {Phys. Rev. A}}\ }%
  \textbf{\bibinfo {volume} {82}},\ \bibinfo {pages} {033609} (\bibinfo {month}
  {Sep}\ \bibinfo {year} {2010})%
  \bibAnnoteFile{NoStop}{PhysRevA.82.033609}%
\bibitem{spinor.rev}%
  \BibitemOpen
  \bibfield{author}{%
  \bibinfo {author} {\bibfnamefont{D.~M.}\ \bibnamefont{Stamper-Kurn}}\ and\
  \bibinfo {author} {\bibfnamefont{M.}~\bibnamefont{Ueda}},\ }%
  \bibfield{journal}{%
  \bibinfo {journal} {arxiv}}%
   (\bibinfo {month} {May}\ \bibinfo {year} {2012}),\ \doi{\bibinfo {doi}
  {arXiv:1205.1888v1}},\ \url{http://arxiv.org/abs/1205.1888}%
  \bibAnnoteFile{NoStop}{spinor.rev}%
\bibitem{PhysRevLett.77.3276}%
  \BibitemOpen
  \bibfield{author}{%
  \bibinfo {author} {\bibfnamefont{T.-L.}\ \bibnamefont{Ho}}\ and\ \bibinfo
  {author} {\bibfnamefont{V.~B.}\ \bibnamefont{Shenoy}},\ }%
  \bibfield{journal}{%
  \Doi{10.1103/PhysRevLett.77.3276}{\bibinfo {journal} {Phys. Rev. Lett.}}\ }%
  \textbf{\bibinfo {volume} {77}},\ \bibinfo {pages} {3276} (\bibinfo {month}
  {Oct}\ \bibinfo {year} {1996})%
  \bibAnnoteFile{NoStop}{PhysRevLett.77.3276}%
\bibitem{PhysRevLett.80.1130}%
  \BibitemOpen
  \bibfield{author}{%
  \bibinfo {author} {\bibfnamefont{H.}~\bibnamefont{Pu}}\ and\ \bibinfo
  {author} {\bibfnamefont{N.~P.}\ \bibnamefont{Bigelow}},\ }%
  \bibfield{journal}{%
  \Doi{10.1103/PhysRevLett.80.1130}{\bibinfo {journal} {Phys. Rev. Lett.}}\ }%
  \textbf{\bibinfo {volume} {80}},\ \bibinfo {pages} {1130} (\bibinfo {month}
  {Feb}\ \bibinfo {year} {1998})%
  \bibAnnoteFile{NoStop}{PhysRevLett.80.1130}%
\bibitem{PhysRevLett.81.5718}%
  \BibitemOpen
  \bibfield{author}{%
  \bibinfo {author} {\bibfnamefont{E.}~\bibnamefont{Timmermans}},\ }%
  \bibfield{journal}{%
  \Doi{10.1103/PhysRevLett.81.5718}{\bibinfo {journal} {Phys. Rev. Lett.}}\ }%
  \textbf{\bibinfo {volume} {81}},\ \bibinfo {pages} {5718} (\bibinfo {month}
  {Dec}\ \bibinfo {year} {1998})%
  \bibAnnoteFile{NoStop}{PhysRevLett.81.5718}%
\bibitem{PhysRevA.58.4836}%
  \BibitemOpen
  \bibfield{author}{%
  \bibinfo {author} {\bibfnamefont{P.}~\bibnamefont{Ao}}\ and\ \bibinfo
  {author} {\bibfnamefont{S.~T.}\ \bibnamefont{Chui}},\ }%
  \bibfield{journal}{%
  \Doi{10.1103/PhysRevA.58.4836}{\bibinfo {journal} {Phys. Rev. A}}\ }%
  \textbf{\bibinfo {volume} {58}},\ \bibinfo {pages} {4836} (\bibinfo {month}
  {Dec}\ \bibinfo {year} {1998})%
  \bibAnnoteFile{NoStop}{PhysRevA.58.4836}%
\bibitem{Tripp}%
  \BibitemOpen
  \bibfield{author}{%
  \bibinfo {author} {\bibfnamefont{M.}~\bibnamefont{﻿Trippenbach}}, \bibinfo
  {author} {\bibfnamefont{K.}~\bibnamefont{Góral}}, \bibinfo {author}
  {\bibfnamefont{K.}~\bibnamefont{Rzazewski}}, \bibinfo {author}
  {\bibfnamefont{B.}~\bibnamefont{Malomed}},\ and\ \bibinfo {author}
  {\bibfnamefont{Y.~B.}\ \bibnamefont{Band}},\ }%
  \bibfield{journal}{%
  \Doi{10.1088/0953-4075/33/19/314}{\bibinfo {journal} {Journal of Physics B:
  Atomic, Molecular and Optical Physics}}\ }%
  \textbf{\bibinfo {volume} {33}},\ \bibinfo {pages} {4017} (\bibinfo {year}
  {2000})%
  \bibAnnoteFile{NoStop}{Tripp}%
\bibitem{PhysRevA.66.013612}%
  \BibitemOpen
  \bibfield{author}{%
  \bibinfo {author} {\bibfnamefont{R.~A.}\ \bibnamefont{Barankov}},\ }%
  \bibfield{journal}{%
  \Doi{10.1103/PhysRevA.66.013612}{\bibinfo {journal} {Phys. Rev. A}}\ }%
  \textbf{\bibinfo {volume} {66}},\ \bibinfo {pages} {013612} (\bibinfo {month}
  {Jul}\ \bibinfo {year} {2002})%
  \bibAnnoteFile{NoStop}{PhysRevA.66.013612}%
\bibitem{PhysRevA.78.023624}%
  \BibitemOpen
  \bibfield{author}{%
  \bibinfo {author} {\bibfnamefont{B.}~\bibnamefont{Van~Schaeybroeck}},\ }%
  \bibfield{journal}{%
  \Doi{10.1103/PhysRevA.78.023624}{\bibinfo {journal} {Phys. Rev. A}}\ }%
  \textbf{\bibinfo {volume} {78}},\ \bibinfo {pages} {023624} (\bibinfo {month}
  {Aug}\ \bibinfo {year} {2008})%
  \bibAnnoteFile{NoStop}{PhysRevA.78.023624}%
\bibitem{0953-4075-43-9-095302}%
  \BibitemOpen
  \bibfield{author}{%
  \bibinfo {author} {\bibfnamefont{S.}~\bibnamefont{Gautam}}\ and\ \bibinfo
  {author} {\bibfnamefont{D.}~\bibnamefont{Angom}},\ }%
  \bibfield{journal}{%
  \bibinfo {journal} {Journal of Physics B: Atomic, Molecular and Optical
  Physics}\ }%
  \textbf{\bibinfo {volume} {43}},\ \bibinfo {pages} {095302} (\bibinfo {year}
  {2010}),\ \url{http://stacks.iop.org/0953-4075/43/i=9/a=095302}%
  \bibAnnoteFile{NoStop}{0953-4075-43-9-095302}%
\bibitem{PhysRevA.66.015602}%
  \BibitemOpen
  \bibfield{author}{%
  \bibinfo {author} {\bibfnamefont{D.~M.}\ \bibnamefont{Jezek}}\ and\ \bibinfo
  {author} {\bibfnamefont{P.}~\bibnamefont{Capuzzi}},\ }%
  \bibfield{journal}{%
  \Doi{10.1103/PhysRevA.66.015602}{\bibinfo {journal} {Phys. Rev. A}}\ }%
  \textbf{\bibinfo {volume} {66}},\ \bibinfo {pages} {015602} (\bibinfo {month}
  {Jul}\ \bibinfo {year} {2002})%
  \bibAnnoteFile{NoStop}{PhysRevA.66.015602}%
\bibitem{PhysRevA.58.1440}%
  \BibitemOpen
  \bibfield{author}{%
  \bibinfo {author} {\bibfnamefont{D.}~\bibnamefont{Gordon}}\ and\ \bibinfo
  {author} {\bibfnamefont{C.~M.}\ \bibnamefont{Savage}},\ }%
  \bibfield{journal}{%
  \Doi{10.1103/PhysRevA.58.1440}{\bibinfo {journal} {Phys. Rev. A}}\ }%
  \textbf{\bibinfo {volume} {58}},\ \bibinfo {pages} {1440} (\bibinfo {month}
  {Aug}\ \bibinfo {year} {1998})%
  \bibAnnoteFile{NoStop}{PhysRevA.58.1440}%
\bibitem{PhysRevA.62.053601}%
  \BibitemOpen
  \bibfield{author}{%
  \bibinfo {author} {\bibfnamefont{B.}~\bibnamefont{Tanatar}}\ and\ \bibinfo
  {author} {\bibfnamefont{K.}~\bibnamefont{Erkan}},\ }%
  \bibfield{journal}{%
  \Doi{10.1103/PhysRevA.62.053601}{\bibinfo {journal} {Phys. Rev. A}}\ }%
  \textbf{\bibinfo {volume} {62}},\ \bibinfo {pages} {053601} (\bibinfo {month}
  {Oct}\ \bibinfo {year} {2000})%
  \bibAnnoteFile{NoStop}{PhysRevA.62.053601}%
\bibitem{PhysRevE.65.066201}%
  \BibitemOpen
  \bibfield{author}{%
  \bibinfo {author} {\bibfnamefont{J.~G.}\ \bibnamefont{Kim}}\ and\ \bibinfo
  {author} {\bibfnamefont{E.~K.}\ \bibnamefont{Lee}},\ }%
  \bibfield{journal}{%
  \Doi{10.1103/PhysRevE.65.066201}{\bibinfo {journal} {Phys. Rev. E}}\ }%
  \textbf{\bibinfo {volume} {65}},\ \bibinfo {pages} {066201} (\bibinfo {month}
  {Jun}\ \bibinfo {year} {2002})%
  \bibAnnoteFile{NoStop}{PhysRevE.65.066201}%
\bibitem{PhysRevA.85.023613}%
  \BibitemOpen
  \bibfield{author}{%
  \bibinfo {author} {\bibfnamefont{A.}~\bibnamefont{Bala\ifmmode~\check{z}\else
  \v{z}\fi{}}}\ and\ \bibinfo {author} {\bibfnamefont{A.~I.}\
  \bibnamefont{Nicolin}},\ }%
  \bibfield{journal}{%
  \Doi{10.1103/PhysRevA.85.023613}{\bibinfo {journal} {Phys. Rev. A}}\ }%
  \textbf{\bibinfo {volume} {85}},\ \bibinfo {pages} {023613} (\bibinfo {month}
  {Feb}\ \bibinfo {year} {2012}),\
  \url{http://link.aps.org/doi/10.1103/PhysRevA.85.023613}%
  \bibAnnoteFile{NoStop}{PhysRevA.85.023613}%
\bibitem{PhysRevA.64.021601}%
  \BibitemOpen
  \bibfield{author}{%
  \bibinfo {author} {\bibfnamefont{N.~P.}\ \bibnamefont{Robins}}, \bibinfo
  {author} {\bibfnamefont{W.}~\bibnamefont{Zhang}}, \bibinfo {author}
  {\bibfnamefont{E.~A.}\ \bibnamefont{Ostrovskaya}},\ and\ \bibinfo {author}
  {\bibfnamefont{Y.~S.}\ \bibnamefont{Kivshar}},\ }%
  \bibfield{journal}{%
  \Doi{10.1103/PhysRevA.64.021601}{\bibinfo {journal} {Phys. Rev. A}}\ }%
  \textbf{\bibinfo {volume} {64}},\ \bibinfo {pages} {021601} (\bibinfo {month}
  {Jul}\ \bibinfo {year} {2001})%
  \bibAnnoteFile{NoStop}{PhysRevA.64.021601}%
\bibitem{Shukla_Sten_Fedele_2001}%
  \BibitemOpen
  \bibfield{author}{%
  \bibinfo {author} {\bibfnamefont{P.~K.}\ \bibnamefont{Shukla}}, \bibinfo
  {author} {\bibfnamefont{L.}~\bibnamefont{Sten}},\ and\ \bibinfo {author}
  {\bibfnamefont{R.}~\bibnamefont{Fedele}},\ }%
  \bibfield{journal}{%
  \bibinfo {journal} {Physica}\ }%
  \textbf{\bibinfo {volume} {553}},\ \bibinfo {pages} {6} (\bibinfo {year}
  {2001})%
  \bibAnnoteFile{NoStop}{Shukla_Sten_Fedele_2001}%
\bibitem{PhysRevLett.93.100402}%
  \BibitemOpen
  \bibfield{author}{%
  \bibinfo {author} {\bibfnamefont{K.}~\bibnamefont{Kasamatsu}}\ and\ \bibinfo
  {author} {\bibfnamefont{M.}~\bibnamefont{Tsubota}},\ }%
  \bibfield{journal}{%
  \Doi{10.1103/PhysRevLett.93.100402}{\bibinfo {journal} {Phys. Rev. Lett.}}\
  }%
  \textbf{\bibinfo {volume} {93}},\ \bibinfo {pages} {100402} (\bibinfo {month}
  {Sep}\ \bibinfo {year} {2004})%
  \bibAnnoteFile{NoStop}{PhysRevLett.93.100402}%
\bibitem{Kourakis_Shukla_Marklund_Stenflo_2005}%
  \BibitemOpen
  \bibfield{author}{%
  \bibinfo {author} {\bibfnamefont{I.}~\bibnamefont{Kourakis}}, \bibinfo
  {author} {\bibfnamefont{P.~K.}\ \bibnamefont{Shukla}}, \bibinfo {author}
  {\bibfnamefont{M.}~\bibnamefont{Marklund}},\ and\ \bibinfo {author}
  {\bibfnamefont{L.}~\bibnamefont{Stenflo}},\ }%
  \bibfield{journal}{%
  \bibinfo {journal} {European Physical Journal B}\ }%
  \textbf{\bibinfo {volume} {46}},\ \bibinfo {pages} {381} (\bibinfo {year}
  {2005})%
  \bibAnnoteFile{NoStop}{Kourakis_Shukla_Marklund_Stenflo_2005}%
\bibitem{PhysRevA.71.035601}%
  \BibitemOpen
  \bibfield{author}{%
  \bibinfo {author} {\bibfnamefont{T.~S.}\ \bibnamefont{Raju}}, \bibinfo
  {author} {\bibfnamefont{P.~K.}\ \bibnamefont{Panigrahi}},\ and\ \bibinfo
  {author} {\bibfnamefont{K.}~\bibnamefont{Porsezian}},\ }%
  \bibfield{journal}{%
  \Doi{10.1103/PhysRevA.71.035601}{\bibinfo {journal} {Phys. Rev. A}}\ }%
  \textbf{\bibinfo {volume} {71}},\ \bibinfo {pages} {035601} (\bibinfo {month}
  {Mar}\ \bibinfo {year} {2005})%
  \bibAnnoteFile{NoStop}{PhysRevA.71.035601}%
\bibitem{modulation}%
  \BibitemOpen
  \bibfield{author}{%
  \bibinfo {author} {\bibfnamefont{S.}~\bibnamefont{Ronen}}, \bibinfo {author}
  {\bibfnamefont{J.~L.}\ \bibnamefont{Bohn}}, \bibinfo {author}
  {\bibfnamefont{L.~E.}\ \bibnamefont{Halmo}},\ and\ \bibinfo {author}
  {\bibfnamefont{M.}~\bibnamefont{Edwards}},\ }%
  \bibfield{journal}{%
  \Doi{10.1103/PhysRevA.78.053613}{\bibinfo {journal} {Phys. Rev. A}}\ }%
  \textbf{\bibinfo {volume} {78}},\ \bibinfo {pages} {053613} (\bibinfo {month}
  {Nov}\ \bibinfo {year} {2008})%
  \bibAnnoteFile{NoStop}{modulation}%
\bibitem{PhysRevLett.87.010401}%
  \BibitemOpen
  \bibfield{author}{%
  \bibinfo {author} {\bibfnamefont{T.}~\bibnamefont{Busch}}\ and\ \bibinfo
  {author} {\bibfnamefont{J.~R.}\ \bibnamefont{Anglin}},\ }%
  \bibfield{journal}{%
  \Doi{10.1103/PhysRevLett.87.010401}{\bibinfo {journal} {Phys. Rev. Lett.}}\
  }%
  \textbf{\bibinfo {volume} {87}},\ \bibinfo {pages} {010401} (\bibinfo {month}
  {Jun}\ \bibinfo {year} {2001})%
  \bibAnnoteFile{NoStop}{PhysRevLett.87.010401}%
\bibitem{1367-2630-5-1-364}%
  \BibitemOpen
  \bibfield{author}{%
  \bibinfo {author} {\bibfnamefont{P.~G.}\ \bibnamefont{Kevrekidis}}, \bibinfo
  {author} {\bibfnamefont{D.~J.}\ \bibnamefont{Frantzeskakis}}, \bibinfo
  {author} {\bibfnamefont{B.~A.}\ \bibnamefont{Malomed}}, \bibinfo {author}
  {\bibfnamefont{A.~R.}\ \bibnamefont{Bishop}},\ and\ \bibinfo {author}
  {\bibfnamefont{I.~G.}\ \bibnamefont{Kevrekidis}},\ }%
  \bibfield{journal}{%
  \bibinfo {journal} {New Journal of Physics}\ }%
  \textbf{\bibinfo {volume} {5}},\ \bibinfo {pages} {64} (\bibinfo {year}
  {2003}),\ \url{http://stacks.iop.org/1367-2630/5/i=1/a=364}%
  \bibAnnoteFile{NoStop}{1367-2630-5-1-364}%
\bibitem{Becker_Stellmer_Soltan-Panahi_Drscher_Baumert_Richter_Kronjger_Bongs_%
Sengstock_2008}%
  \BibitemOpen
  \bibfield{author}{%
  \bibinfo {author} {\bibfnamefont{C.}~\bibnamefont{Becker}}, \bibinfo {author}
  {\bibfnamefont{S.}~\bibnamefont{Stellmer}}, \bibinfo {author}
  {\bibfnamefont{P.}~\bibnamefont{Soltan-Panahi}}, \bibinfo {author}
  {\bibfnamefont{S.}~\bibnamefont{D{\"o}rscher}}, \bibinfo {author}
  {\bibfnamefont{M.}~\bibnamefont{Baumert}}, \bibinfo {author}
  {\bibfnamefont{E.~M.}\ \bibnamefont{Richter}}, \bibinfo {author}
  {\bibfnamefont{J.}~\bibnamefont{Kronj{\"a}ger}}, \bibinfo {author}
  {\bibfnamefont{K.}~\bibnamefont{Bongs}},\ and\ \bibinfo {author}
  {\bibfnamefont{K.}~\bibnamefont{Sengstock}},\ }%
  \bibfield{journal}{%
  \bibinfo {journal} {Nature Physics}\ }%
  \textbf{\bibinfo {volume} {4}},\ \bibinfo {pages} {9} (\bibinfo {year}
  {2008})%
  \bibAnnoteFile{NoStop}{Becker_Stellmer_Soltan-Panahi_Drscher_Baumert_Richter%
_Kronjger_Bongs_Sengstock_2008}%
\bibitem{PhysRevLett.106.065302}%
  \BibitemOpen
  \bibfield{author}{%
  \bibinfo {author} {\bibfnamefont{C.}~\bibnamefont{Hamner}}, \bibinfo {author}
  {\bibfnamefont{J.~J.}\ \bibnamefont{Chang}}, \bibinfo {author}
  {\bibfnamefont{P.}~\bibnamefont{Engels}},\ and\ \bibinfo {author}
  {\bibfnamefont{M.~A.}\ \bibnamefont{Hoefer}},\ }%
  \bibfield{journal}{%
  \Doi{10.1103/PhysRevLett.106.065302}{\bibinfo {journal} {Phys. Rev. Lett.}}\
  }%
  \textbf{\bibinfo {volume} {106}},\ \bibinfo {pages} {065302} (\bibinfo
  {month} {Feb}\ \bibinfo {year} {2011}),\
  \url{http://link.aps.org/doi/10.1103/PhysRevLett.106.065302}%
  \bibAnnoteFile{NoStop}{PhysRevLett.106.065302}%
\bibitem{PhysRevLett.82.4956}%
  \BibitemOpen
  \bibfield{author}{%
  \bibinfo {author} {\bibfnamefont{D.~L.}\ \bibnamefont{Feder}}, \bibinfo
  {author} {\bibfnamefont{C.~W.}\ \bibnamefont{Clark}},\ and\ \bibinfo {author}
  {\bibfnamefont{B.~I.}\ \bibnamefont{Schneider}},\ }%
  \bibfield{journal}{%
  \Doi{10.1103/PhysRevLett.82.4956}{\bibinfo {journal} {Phys. Rev. Lett.}}\ }%
  \textbf{\bibinfo {volume} {82}},\ \bibinfo {pages} {4956} (\bibinfo {month}
  {Jun}\ \bibinfo {year} {1999})%
  \bibAnnoteFile{NoStop}{PhysRevLett.82.4956}%
\bibitem{PhysRevLett.78.3594}%
  \BibitemOpen
  \bibfield{author}{%
  \bibinfo {author} {\bibfnamefont{B.~D.}\ \bibnamefont{Esry}}, \bibinfo
  {author} {\bibfnamefont{C.~H.}\ \bibnamefont{Greene}}, \bibinfo {author}
  {\bibfnamefont{J.~P.}\ \bibnamefont{Burke}, \bibfnamefont{Jr.}},\ and\
  \bibinfo {author} {\bibfnamefont{J.~L.}\ \bibnamefont{Bohn}},\ }%
  \bibfield{journal}{%
  \Doi{10.1103/PhysRevLett.78.3594}{\bibinfo {journal} {Phys. Rev. Lett.}}\ }%
  \textbf{\bibinfo {volume} {78}},\ \bibinfo {pages} {3594} (\bibinfo {month}
  {May}\ \bibinfo {year} {1997}),\
  \url{http://link.aps.org/doi/10.1103/PhysRevLett.78.3594}%
  \bibAnnoteFile{NoStop}{PhysRevLett.78.3594}%
\bibitem{PhysRevA.57.1272}%
  \BibitemOpen
  \bibfield{author}{%
  \bibinfo {author} {\bibfnamefont{P.}~\bibnamefont{\"Ohberg}}\ and\ \bibinfo
  {author} {\bibfnamefont{S.}~\bibnamefont{Stenholm}},\ }%
  \bibfield{journal}{%
  \Doi{10.1103/PhysRevA.57.1272}{\bibinfo {journal} {Phys. Rev. A}}\ }%
  \textbf{\bibinfo {volume} {57}},\ \bibinfo {pages} {1272} (\bibinfo {month}
  {Feb}\ \bibinfo {year} {1998}),\
  \url{http://link.aps.org/doi/10.1103/PhysRevA.57.1272}%
  \bibAnnoteFile{NoStop}{PhysRevA.57.1272}%
\bibitem{PhysRevA.75.013601}%
  \BibitemOpen
  \bibfield{author}{%
  \bibinfo {author} {\bibfnamefont{C.-H.}\ \bibnamefont{Zhang}}\ and\ \bibinfo
  {author} {\bibfnamefont{H.~A.}\ \bibnamefont{Fertig}},\ }%
  \bibfield{journal}{%
  \Doi{10.1103/PhysRevA.75.013601}{\bibinfo {journal} {Phys. Rev. A}}\ }%
  \textbf{\bibinfo {volume} {75}},\ \bibinfo {pages} {013601} (\bibinfo {month}
  {Jan}\ \bibinfo {year} {2007}),\
  \url{http://link.aps.org/doi/10.1103/PhysRevA.75.013601}%
  \bibAnnoteFile{NoStop}{PhysRevA.75.013601}%
\bibitem{PhysRevA.77.033606}%
  \BibitemOpen
  \bibfield{author}{%
  \bibinfo {author} {\bibfnamefont{M.~O.~C.}\ \bibnamefont{Pires}}\ and\
  \bibinfo {author} {\bibfnamefont{E.~J.~V.}\ \bibnamefont{de~Passos}},\ }%
  \bibfield{journal}{%
  \Doi{10.1103/PhysRevA.77.033606}{\bibinfo {journal} {Phys. Rev. A}}\ }%
  \textbf{\bibinfo {volume} {77}},\ \bibinfo {pages} {033606} (\bibinfo {month}
  {Mar}\ \bibinfo {year} {2008}),\
  \url{http://link.aps.org/doi/10.1103/PhysRevA.77.033606}%
  \bibAnnoteFile{NoStop}{PhysRevA.77.033606}%
\bibitem{PhysRevA.70.063606}%
  \BibitemOpen
  \bibfield{author}{%
  \bibinfo {author} {\bibfnamefont{H.}~\bibnamefont{Ma}}\ and\ \bibinfo
  {author} {\bibfnamefont{T.}~\bibnamefont{Pang}},\ }%
  \bibfield{journal}{%
  \Doi{10.1103/PhysRevA.70.063606}{\bibinfo {journal} {Phys. Rev. A}}\ }%
  \textbf{\bibinfo {volume} {70}},\ \bibinfo {pages} {063606} (\bibinfo {month}
  {Dec}\ \bibinfo {year} {2004})%
  \bibAnnoteFile{NoStop}{PhysRevA.70.063606}%
\bibitem{RevModPhys.71.463}%
  \BibitemOpen
  \bibfield{author}{%
  \bibinfo {author} {\bibfnamefont{F.}~\bibnamefont{Dalfovo}}, \bibinfo
  {author} {\bibfnamefont{S.}~\bibnamefont{Giorgini}}, \bibinfo {author}
  {\bibfnamefont{L.~P.}\ \bibnamefont{Pitaevskii}},\ and\ \bibinfo {author}
  {\bibfnamefont{S.}~\bibnamefont{Stringari}},\ }%
  \bibfield{journal}{%
  \Doi{10.1103/RevModPhys.71.463}{\bibinfo {journal} {Rev. Mod. Phys.}}\ }%
  \textbf{\bibinfo {volume} {71}},\ \bibinfo {pages} {463} (\bibinfo {month}
  {Apr}\ \bibinfo {year} {1999})%
  \bibAnnoteFile{NoStop}{RevModPhys.71.463}%
\bibitem{Pethick2002}%
  \BibitemOpen
  \bibfield{author}{%
  \bibinfo {author} {\bibfnamefont{C.}~\bibnamefont{Pethick}}\ and\ \bibinfo
  {author} {\bibfnamefont{H.}~\bibnamefont{Smith}},\ }%
  \emph{\bibinfo {title} {Bose-Einstein condensation in dilute gases}}\
  (\bibinfo {publisher} {Cambridge University Press},\ \bibinfo {year} {2002})%
  \bibAnnoteFile{NoStop}{Pethick2002}%
\bibitem{PhysRevA.85.043602}%
  \BibitemOpen
  \bibfield{author}{%
  \bibinfo {author} {\bibfnamefont{L.}~\bibnamefont{Wen}}, \bibinfo {author}
  {\bibfnamefont{W.~M.}\ \bibnamefont{Liu}}, \bibinfo {author}
  {\bibfnamefont{Y.}~\bibnamefont{Cai}}, \bibinfo {author}
  {\bibfnamefont{J.~M.}\ \bibnamefont{Zhang}},\ and\ \bibinfo {author}
  {\bibfnamefont{J.}~\bibnamefont{Hu}},\ }%
  \bibfield{journal}{%
  \Doi{10.1103/PhysRevA.85.043602}{\bibinfo {journal} {Phys. Rev. A}}\ }%
  \textbf{\bibinfo {volume} {85}},\ \bibinfo {pages} {043602} (\bibinfo {month}
  {Apr}\ \bibinfo {year} {2012})%
  \bibAnnoteFile{NoStop}{PhysRevA.85.043602}%
\bibitem{PhysRevA.53.2477}%
  \BibitemOpen
  \bibfield{author}{%
  \bibinfo {author} {\bibfnamefont{F.}~\bibnamefont{Dalfovo}}\ and\ \bibinfo
  {author} {\bibfnamefont{S.}~\bibnamefont{Stringari}},\ }%
  \bibfield{journal}{%
  \Doi{10.1103/PhysRevA.53.2477}{\bibinfo {journal} {Phys. Rev. A}}\ }%
  \textbf{\bibinfo {volume} {53}},\ \bibinfo {pages} {2477} (\bibinfo {month}
  {Apr}\ \bibinfo {year} {1996})%
  \bibAnnoteFile{NoStop}{PhysRevA.53.2477}%
\bibitem{jenkin_2011}%
  \BibitemOpen
  \bibfield{author}{%
  \bibinfo {author} {\bibfnamefont{D.}~\bibnamefont{Gordon}}\ and\ \bibinfo
  {author} {\bibfnamefont{C.~M.}\ \bibnamefont{Savage}},\ }%
  \bibfield{journal}{%
  \Doi{10.1103/PhysRevA.58.1440}{\bibinfo {journal} {Phys. Rev. A}}\ }%
  \textbf{\bibinfo {volume} {58}},\ \bibinfo {pages} {1440} (\bibinfo {month}
  {Aug}\ \bibinfo {year} {1998}),\
  \url{http://link.aps.org/doi/10.1103/PhysRevA.58.1440}%
  \bibAnnoteFile{NoStop}{jenkin_2011}%
\bibitem{cho_2011}%
  \BibitemOpen
  \bibfield{author}{%
  \bibinfo {author} {\bibfnamefont{D.}~\bibnamefont{Jenkin}}, \bibinfo {author}
  {\bibfnamefont{D.}~\bibnamefont{McCarron}}, \bibinfo {author}
  {\bibfnamefont{M.}~\bibnamefont{Köppinger}}, \bibinfo {author}
  {\bibfnamefont{H.}~\bibnamefont{Cho}}, \bibinfo {author}
  {\bibfnamefont{S.}~\bibnamefont{Hopkins}},\ and\ \bibinfo {author}
  {\bibfnamefont{S.}~\bibnamefont{Cornish}},\ }%
  \bibfield{journal}{%
  \bibinfo {journal} {The European Physical Journal D - Atomic, Molecular,
  Optical and Plasma Physics}\ }%
  \textbf{\bibinfo {volume} {65}},\ \bibinfo {pages} {11} (\bibinfo {year}
  {2011}),\ ISSN \bibinfo {issn} {1434-6060},\ \bibinfo {note}
  {10.1140/epjd/e2011-10720-5},\
  \url{http://dx.doi.org/10.1140/epjd/e2011-10720-5}%
  \bibAnnoteFile{NoStop}{cho_2011}%
\bibitem{exp_note}%
  \BibitemOpen
  \bibfield{author}{%
  \bibinfo {author} {\bibfnamefont{H.}~\bibnamefont{Cho}}, \bibinfo {author}
  {\bibfnamefont{D.}~\bibnamefont{McCarron}}, \bibinfo {author}
  {\bibfnamefont{D.}~\bibnamefont{Jenkin}}, \bibinfo {author}
  {\bibfnamefont{M.}~\bibnamefont{Köppinger}},\ and\ \bibinfo {author}
  {\bibfnamefont{S.}~\bibnamefont{Cornish}},\ }%
  \bibfield{journal}{%
  \bibinfo {journal} {The European Physical Journal D - Atomic, Molecular,
  Optical and Plasma Physics}\ }%
  \textbf{\bibinfo {volume} {65}},\ \bibinfo {pages} {125} (\bibinfo {year}
  {2011}),\ ISSN \bibinfo {issn} {1434-6060},\ \bibinfo {note}
  {10.1140/epjd/e2011-10716-1},\
  \url{http://dx.doi.org/10.1140/epjd/e2011-10716-1}%
  \bibAnnoteFile{NoStop}{exp_note}%
\bibitem{PhysRevLett.89.283202}%
  \BibitemOpen
  \bibfield{author}{%
  \bibinfo {author} {\bibfnamefont{A.}~\bibnamefont{Marte}}, \bibinfo {author}
  {\bibfnamefont{T.}~\bibnamefont{Volz}}, \bibinfo {author}
  {\bibfnamefont{J.}~\bibnamefont{Schuster}}, \bibinfo {author}
  {\bibfnamefont{S.}~\bibnamefont{D\"urr}}, \bibinfo {author}
  {\bibfnamefont{G.}~\bibnamefont{Rempe}}, \bibinfo {author}
  {\bibfnamefont{E.~G.~M.}\ \bibnamefont{van Kempen}},\ and\ \bibinfo {author}
  {\bibfnamefont{B.~J.}\ \bibnamefont{Verhaar}},\ }%
  \bibfield{journal}{%
  \Doi{10.1103/PhysRevLett.89.283202}{\bibinfo {journal} {Phys. Rev. Lett.}}\
  }%
  \textbf{\bibinfo {volume} {89}},\ \bibinfo {pages} {283202} (\bibinfo {month}
  {Dec}\ \bibinfo {year} {2002}),\
  \url{http://link.aps.org/doi/10.1103/PhysRevLett.89.283202}%
  \bibAnnoteFile{NoStop}{PhysRevLett.89.283202}%
\bibitem{PhysRevA.70.032701}%
  \BibitemOpen
  \bibfield{author}{%
  \bibinfo {author} {\bibfnamefont{C.}~\bibnamefont{Chin}}, \bibinfo {author}
  {\bibfnamefont{V.}~\bibnamefont{Vuleti\ifmmode~\acute{c}\else \'{c}\fi{}}},
  \bibinfo {author} {\bibfnamefont{A.~J.}\ \bibnamefont{Kerman}}, \bibinfo
  {author} {\bibfnamefont{S.}~\bibnamefont{Chu}}, \bibinfo {author}
  {\bibfnamefont{E.}~\bibnamefont{Tiesinga}}, \bibinfo {author}
  {\bibfnamefont{P.~J.}\ \bibnamefont{Leo}},\ and\ \bibinfo {author}
  {\bibfnamefont{C.~J.}\ \bibnamefont{Williams}},\ }%
  \bibfield{journal}{%
  \Doi{10.1103/PhysRevA.70.032701}{\bibinfo {journal} {Phys. Rev. A}}\ }%
  \textbf{\bibinfo {volume} {70}},\ \bibinfo {pages} {032701} (\bibinfo {month}
  {Sep}\ \bibinfo {year} {2004}),\
  \url{http://link.aps.org/doi/10.1103/PhysRevA.70.032701}%
  \bibAnnoteFile{NoStop}{PhysRevA.70.032701}%
\bibitem{PhysRevA.85.032506}%
  \BibitemOpen
  \bibfield{author}{%
  \bibinfo {author} {\bibfnamefont{T.}~\bibnamefont{Takekoshi}}, \bibinfo
  {author} {\bibfnamefont{M.}~\bibnamefont{Debatin}}, \bibinfo {author}
  {\bibfnamefont{R.}~\bibnamefont{Rameshan}}, \bibinfo {author}
  {\bibfnamefont{F.}~\bibnamefont{Ferlaino}}, \bibinfo {author}
  {\bibfnamefont{R.}~\bibnamefont{Grimm}}, \bibinfo {author}
  {\bibfnamefont{H.-C.}\ \bibnamefont{N\"agerl}}, \bibinfo {author}
  {\bibfnamefont{C.~R.}\ \bibnamefont{Le~Sueur}}, \bibinfo {author}
  {\bibfnamefont{J.~M.}\ \bibnamefont{Hutson}}, \bibinfo {author}
  {\bibfnamefont{P.~S.}\ \bibnamefont{Julienne}}, \bibinfo {author}
  {\bibfnamefont{S.}~\bibnamefont{Kotochigova}},\ and\ \bibinfo {author}
  {\bibfnamefont{E.}~\bibnamefont{Tiemann}},\ }%
  \bibfield{journal}{%
  \Doi{10.1103/PhysRevA.85.032506}{\bibinfo {journal} {Phys. Rev. A}}\ }%
  \textbf{\bibinfo {volume} {85}},\ \bibinfo {pages} {032506} (\bibinfo {month}
  {Mar}\ \bibinfo {year} {2012}),\
  \url{http://link.aps.org/doi/10.1103/PhysRevA.85.032506}%
  \bibAnnoteFile{NoStop}{PhysRevA.85.032506}%
\bibitem{post}%
  \BibitemOpen
  \bibfield{author}{%
  \bibinfo {author} {\bibfnamefont{I.-K.}\ \bibnamefont{Liu}}, \bibinfo
  {author} {\bibfnamefont{H.~W.}\ \bibnamefont{Cho}}, \bibinfo {author}
  {\bibfnamefont{S.-W.}\ \bibnamefont{Su}}, \bibinfo {author}
  {\bibfnamefont{D.~J.}\ \bibnamefont{McCarron}}, \bibinfo {author}
  {\bibfnamefont{S.~L.}\ \bibnamefont{Cornish}},\ and\ \bibinfo {author}
  {\bibfnamefont{S.-C.}\ \bibnamefont{Gou}},\ }%
  \bibinfo {howpublished} {Poster presented at Internation Conference on Atomic
  Physics (ICAP) 2012} (\bibinfo {year} {2012})%
  \bibAnnoteFile{NoStop}{post}%
\end{thebibliography}
\end{document}